\documentclass[a4paper,amsmath,amssymb,color,nofootinbib,preprint,tightenlines]{revtex4-2}
\usepackage{graphicx}
\usepackage[dvipsnames]{xcolor}
\usepackage[mathscr]{euscript}
\usepackage{bm,multirow,slashed}
\usepackage{tablefootnote}
\usepackage[colorlinks=true,pdfstartview=FitV,citecolor=blue,linkcolor=purple,urlcolor=Sepia]{hyperref}
\usepackage{orcidlink}

\setcitestyle{citesep={,\kern-0.2em}}

\begin{document}
\baselineskip=17pt \parskip=3pt

\title{Probing \boldmath$CP$ violation in weak radiative and dielectron\\decays of $\Lambda$ and $\Xi^-$ hyperons}

\author{Xiao-Gang He\,\orcidlink{0000-0001-7059-6311}$,^{1,2,}$\footnote{\email{hexg@sjtu.edu.cn}} Jusak Tandean\,\orcidlink{0000-0001-7581-2351}$,^{1,2,}$\footnote{\email{jtandean@yahoo.com}} and German Valencia\,\orcidlink{0000-0001-6600-1290}$^{3,}$\footnote{\email{german.valencia@monash.edu}}\vspace{1ex}\\ \it
$^1$State Key Laboratory of Dark Matter Physics,\\ Tsung-Dao Lee Institute \& School of Physics and Astronomy, Shanghai Jiao Tong University, Shanghai 201210, China\vspace{1ex}\\
$^2$Key Laboratory for Particle Astrophysics and Cosmology (MOE)\\ \& Shanghai Key Laboratory for Particle Physics and Cosmology,\\ Tsung-Dao Lee Institute \& School of
Physics and Astronomy, Shanghai Jiao Tong University, Shanghai 201210, China\vspace{1ex}\\
$^3$School of Physics and Astronomy, Monash University, Wellington Road, Clayton, Victoria 3800, Australia\vspace{2em}\\
\rm Abstract \vspace{8pt} \\
\begin{minipage}{0.99\textwidth} \baselineskip=17pt \parindent=3ex \small
The rare decays $\Lambda\to ne^+e^-$ and $\Xi^-\to\Sigma^-e^+e^-$ can serve as sensitive probes of $CP$ violation, especially that induced by interactions from beyond the standard model.
We demonstrate that this is so because the dominant long-distance components of these processes provide sizable absorptive phases that can give rise to significant $CP$ violation via interference with potential new-physics effects.
Treating the new contributions model-independently, we find that this leads to $CP$-odd rate asymmetries that are currently allowed by kaon and hyperon data to reach the 10\% and 20\% levels in the $\Lambda$ and $\Xi^-$ cases, respectively.
Interestingly, we obtain comparable outcomes for the related weak radiative modes $\Lambda\to n\gamma$ and $\Xi^-\to\Sigma^-\gamma$.
These results are testable at future high-statistics hyperon facilities.
\end{minipage}}

\maketitle

\newpage

{\bf\small\hypersetup{linkcolor=black}\tableofcontents}

\newpage

\section{Introduction\label{intro}}

The past couple of decades have witnessed a marked increase in experimental \cite{HyperCP:2005mvo,HyperCP:2005jyj,Li:2016tlt,LHCb:2017rdd,AlvesJunior:2018ldo,Goudzovski:2022vbt,Schonning:2023mge,BESIII:2023utd,BESIII:2024wvw,LHCb:2025evf,Zhang:2026pyv,BESIII:2026ztz} and theoretical \cite{Hu:2018luj,Tandean:2019tkm,Li:2019cbk,Su:2019ipw,MartinCamalich:2020dfe,Wang:2021uzi,Geng:2021fog,He:2023cqg,Roy:2024hqg,Eguren:2024oov,MartinCamalich:2025srw,He:2026chn,Ziegler:2026kis} attention given to rare hyperon processes as an environment in which to search for new physics (NP) beyond the standard model (SM).
This has to do with their being heavily suppressed  or strictly forbidden within the SM, implying that NP might generate sufficiently enhancing effects for them to be discoverable by ongoing or near-future efforts.

We have recently explored this possibility in ref.\,\cite{He:2026chn}, particularly investigating $CP$ violation in \,$\Sigma^+\to p\ell^+\ell^-$, \,$\ell=e,\mu$,\, which is predicted to be small in the SM but which can be much enlarged in the presence of NP.
As we elaborated therein, the dominant photon-mediated long-distance (LD) portions of the amplitudes for these hyperon decays supply important absorptive phases that could drive substantial $CP$ violation through interference with the NP contributions.
Here we extend the study to the $\Lambda$ and $\Xi^-$ hyperons, namely the dielectron channels \,$\Lambda\to ne^+e^-$\, and \,$\Xi^-\to\Sigma^-e^+e^-$,\, which have not yet been measured, as well as the weak radiative ones \,$\Lambda\to n\gamma$\, and \,$\Xi^-\to\Sigma^-\gamma$,\, the dimuon ones being kinematically disallowed.
It turns out that these modes are more advantageous than the $\Sigma^+$ ones in that the $CP$ asymmetries in the former can be amplified over those in the latter, as detailed later on.
It is remarkable that, though the dielectron channels are overwhelmingly controlled by their LD components, these asymmetries are sensitive to the short-distance (SD) amplitudes, which could be influenced by interactions from beyond the SM.
Similar statements can be made of the weak radiative modes.
These findings offer novel ways to look for $CP$ violation in the hyperon sector.

The structure of the rest of the paper is as follows.
In section \ref{ld+sd} we describe the LD and SD contributions to \,$\Lambda\to ne^+e^-,n\gamma$\, and \,$\Xi^-\to\Sigma^-e^+e^-,\Sigma^-\gamma$.\,
Adopting a model-independent approach, we include effective low-energy quark-level operators that can be brought about by NP and modify the SM amplitudes.
In section \ref{cpv-sm-np} we first specify the $CP$-odd observables to be considered and subsequently examine them numerically within the SM and if NP affects them.
We make concluding comments in section \ref{concl}.
In a few appendices we outline the derivation of the LD amplitudes.

\section{Long- and short-distance contributions\label{ld+sd}}

The strangeness changing ($|\Delta S|=1)$ decay \,{\small$\mathfrak B$}$\,\to\,${\small$\mathfrak B$}$'\ell^+\ell^-$\, of a spin-1/2 hyperon {\small$\mathfrak B$} into another spin-1/2 baryon {\small$\mathfrak B$}$'$ and a charged-lepton pair $\ell^+\ell^-$ is dominated by LD contributions proceeding from \,{\small$\mathfrak B$}$\,\to\,${\small$\mathfrak B$}$'\gamma^*$\, with a virtual photon $\gamma^*$.
The amplitude for this photonic transition is expressible as \cite{Lyagin:1962,Bergstrom:1987wr,He:2005yn}
\begin{equation} \label{B2B'g}
{\cal M}_{\mathfrak B\to\mathfrak B'\gamma^*}^{} \,= {\tt\,e}G_{\rm F\,}^{}\bar u_{\mathfrak B'}^{}
\big[ i\sigma^{\nu\kappa} q_\kappa^{} \big(a_{\mathfrak B}^{}+\gamma_5^{}b_{\mathfrak B}^{}\big)
+ \big(q^\nu\slashed q-\gamma^\nu q^2\big)\big(c_{\mathfrak B}^{}+\gamma_5^{}d_{\mathfrak B}^{}\big)
\big] u_{\mathfrak B}^{}\;\! \varepsilon_\nu^* \,, ~~~
\end{equation}
where {\tt e} and $G_{\rm F}$ denote, respectively, the proton's electric charge and the Fermi coupling constant, $\bar u_{\mathfrak B'}^{}$ and $u_{\mathfrak B}^{}$ represent the Dirac spinors of the baryons, \,$q= P_{\mathfrak B}^{}-P_{\mathfrak B'}^{}$\, stands for the difference between their four-momenta, $\varepsilon$ symbolizes the photon's polarization four-vector, \,$\sigma^{\nu\kappa}=i[\gamma^\nu,\gamma^\kappa]/2$,  and $a_{\mathfrak B}^{},\,b_{\mathfrak B}^{},\,c_{\mathfrak B}^{}$, and $d_{\mathfrak B}^{}$ are $q^2$-dependent form-factors which are in general complex.
The imaginary parts of these form factors are reliably calculable based on unitarity arguments~\cite{Farrar:1971xi,Kogan:1983ic,Neufeld:1992np,Jenkins:1992ab} if \,$\mathfrak B'\mathfrak B=n\Lambda,\Sigma^-\Xi^-,p\Sigma^+$\, but vanish if
\,$\mathfrak B'\mathfrak B=\Lambda\Xi^0,\Sigma^0\Xi^0$\, at the lowest loop order in baryon chiral perturbation theory (B$\chi$PT).
In the first three cases, the determination of the imaginary parts as functions of $q^2$ is  sketched in appendices
\ref{L2ng-ff}-\ref{S2pg-ff}.
The $c_{\mathfrak B}^{}$ and $d_{\mathfrak B}^{}$ terms are absent from eq.\,(\ref{B2B'g}) when the photon is real.

For comparison with earlier outcomes of B$\chi$PT, we collect our evaluation of Im\,$a_{\mathfrak B}$ and Im\,$b_{\mathfrak B}$ at \,$q^2=0$\, in columns 2 and 6 of table \ref{imaimb}, where the \,$\Sigma^+{\to}p\gamma$\, entries were previously reported in ref.\,\cite{He:2026chn}.
Columns 3-5 and 7-9 contain the results of refs.\,\cite{Shi:2022dhw,Neufeld:1992np,Jenkins:1992ab}.
Our predictions can be seen to be in good agreement with those of so-called covariant B$\chi$PT in the very recent ref.\,\cite{Shi:2022dhw} and somewhat less so with the ones in ref.\,\cite{Neufeld:1992np} partly due to differences in the input parameters.

\begin{table}[b!]\bigskip\centering\setlength{\tabcolsep}{0ex}\footnotesize
\begin{tabular}{|c|rrcr|rrcr|}\hline
\multirow{2}{*}{\footnotesize Decay} & \multicolumn{4}{c|}{Im\,$a(0)\vphantom{|_|^|}$/MeV} &
\multicolumn{4}{c|}{Im\,$b(0)$/MeV} \\ \cline{2-9} &
\scriptsize\sf Our work ~~ &
\scriptsize\tt Ref$\!$.\cite{Shi:2022dhw} ~ & ~ \scriptsize\tt Ref$\!$.\cite{Neufeld:1992np}~ &  \scriptsize\tt Ref$\!$.\cite{Jenkins:1992ab}~ & \scriptsize\sf Our work ~~ &
\scriptsize\tt Ref$\!$.\cite{Shi:2022dhw} ~ & ~ \scriptsize\tt Ref$\!$.\cite{Neufeld:1992np}~ &  \scriptsize\tt Ref$\!$.\cite{Jenkins:1992ab}~
\\ \hline\hline
\scriptsize$\Lambda\to n\gamma$ & ~$-1.03\pm0.02$ & ~~ $-1.01\pm0.02$ & $-$0.82 & $-$0.68 ~~ & $9.47\pm 0.14\vphantom{|^{|_|^|}}$ & $10.04\pm0.81$ & $~~$9.1 & 11.21 ~~
\\
\scriptsize$~\Xi^-{\to}\;\!\Sigma^-\gamma~$ & $-0.59\pm 0.01$ & $-0.57\pm0.01$ & $-0.64$ & $-$1.55 ~~ & ~$-8.87\pm 0.14\vphantom{|_|^{|^|}}$ & ~~ $-8.64\pm 0.54$ & $-$8.1 & $-$12.34 ~~
\\
\scriptsize$\Sigma^+{\to}\;\!p\gamma$ & $2.95\pm 0.04$ & $2.70\pm 0.04$ & $~$2.6 & 6.18 ~~ & $-1.75\pm 0.03\vphantom{|_|^{|^|}}$ & $-1.75\pm 0.12$ & $-$1.5 & $-$0.53 ~~ \\ \hline
\end{tabular}\caption{The second and sixth columns exhibit our estimation of Im\,$a_{\mathfrak B}$ and Im\,$b_{\mathfrak B}$ at \,$q^2=0$  from unitarity, compared to the results of extended-on-mass-shell covariant B$\chi$PT up to one-loop order in ref.\,\cite{Shi:2022dhw}, of using unitarity plus covariant B$\chi$PT in ref.\,\cite{Neufeld:1992np}, and of heavy-baryon $\chi$PT up to one-loop order in ref.\,\cite{Jenkins:1992ab}.} \label{imaimb}
\end{table}

\begin{table}[b!]\bigskip\centering
\begin{tabular}{|c|cc|}\hline
\footnotesize Decay & ~ $10^4\rm\,BF_{\rm\!absorptive}^{}\vphantom{|_{|_|}^|}$ ~ & $10^4\rm\,BF_{\tt\!exp}^{}$ \\
\hline\hline \footnotesize$\Lambda\to n\gamma$ & $6.1\pm 0.2$ & $8.35\pm 0.65\vphantom{|^{|_o}}$
\\
\footnotesize$~\Xi^-{\to}\;\!\Sigma^-\gamma~$ & $1.30\pm 0.04$ & $1.27\pm 0.23\vphantom{|^|}$
\\
\footnotesize$\Sigma^+{\to}\;\!p\gamma$ & ~~$0.64\pm 0.02$~~ & ~~$9.96\pm 0.28\vphantom{|_o^|}$~~ \\ \hline
\end{tabular}\caption{The second column lists the branching fractions (BFs) calculated solely from our Im\,$a(0)$ and Im\,$b(0)$ numbers in
table\,\,\ref{imaimb} and the third column the observed BFs \cite{BESIII:2022rgl,BGHLQR:1986kqv,E761:1993unn,BESIII:2023fhs,ParticleDataGroup:2026mpi}.} \label{BF}
\end{table}

To look at additional consequences, in the second column of table \ref{BF} we gather the branching fractions of \,{\small$\mathfrak B$}$\,\to\,${\small$\mathfrak B$}$'\gamma$\, from only our Im\,$a_{\mathfrak B}^{}(0)$ and Im\,$b_{\mathfrak B}^{}(0)$ results in table \ref{imaimb}, after employing the rate formula in eq.\,(\ref{B2B'g-rate}) with the SD terms neglected.
The third column of table~\ref{BF} displays their measurements.
It is evident from this table that the absorptive contribution to~\,$\Xi^-{\to}\;\!\Sigma^-\gamma$,\, exclusively from Im\,$a_\Xi^{}(0)$ and Im\,$b_\Xi^{}(0)$, already saturates the observed branching fractions within the errors, and so more precise data would be needed to test our computation more definitively.
We further learn from these two tables that heavy-baryon $\chi$PT up to one-loop order appears to be inadequate for dealing with \,{\small$\mathfrak B$}$\,\to\,${\small$\mathfrak B$}$'\gamma$\, because its implied prediction for the branching fraction of \,$\Xi^-{\to}\;\!\Sigma^-\gamma$\, well exceeds the empirical value.
Another salient conclusion we can draw is that for \,$\Lambda\to n\gamma$\, or
\,$\Xi^-\to\Sigma^-\gamma$\, the absorptive phase of the $b_{\mathfrak B}$ amplitude is several times those of the $a_{\mathfrak B}$ and $b_{\mathfrak B}$ amplitudes for~\,$\Sigma^+\to p\gamma$.\,
This proves to be one of the main factors responsible for the associated $CP$ asymmetries in \,$\Lambda\to n\gamma,ne^+e^-$\, or \,$\Xi^-\to\Sigma^-\gamma,\Sigma^-e^+e^-$\, potentially being bigger than their $\Sigma^+$ counterparts by substantial amounts.
This is illustrated numerically in the next section.

The SD contributions to \,{\small$\mathfrak B$}$\,\to\,${\small$\mathfrak B$}$'e^+e^-$\, arise from the SM and possible NP beyond it.
The effective Hamiltonian for the former is \cite{Shifman:1976de,Buchalla:1995vs,Gilman:1978bg}
\begin{equation} \label{smsd}
{\cal H}_{\rm eff}^{\tt SM} \,=\, \frac{G_{\rm F}}{\sqrt2} \Big[ \big(V_{ud}^*V_{us}^{}z_{7V}^{}
- V_{td}^*V_{ts}^{} y_{7V}^{}\big) O_{7V}^{} - V_{td}^*V_{ts}^{} y_{7A}^{} O_{7A}^{}
+ V_{ud}^*V_{us}^{} C_{7\gamma}^{} O_{7\gamma}^{} \Big] \,, ~~~
\end{equation}
where $V_{kl}$ are elements of the Cabibbo-Kobayashi-Maskawa (CKM) matrix, $z_{7V}^{}$, $y_{7V}^{}$, $y_{7A}^{}$, and $C_{7\gamma}$ denote the Wilson coefficients at QCD renormalization scales of order 1 GeV, and the quark-level operators are
\begin{align} \label{smO}
O_{7V}^{} & \,=\, 2\, \overline d\gamma^\kappa\!P\!_L^{}s\, \overline e\gamma_\kappa^{}e \,, \hspace{5em}
O_{7A}^{} \,=\, 2\, \overline d\gamma^\kappa\!P\!_L^{}s\, \overline e\gamma_\kappa^{}\gamma_5^{}e \,, &
\nonumber \\
O_{7\gamma}^{} & \,=\, \frac{{\tt e} F_{\kappa\nu}}{8\pi^2}\, \overline d\sigma^{\kappa\nu} \big( m_s^{} P\!_R^{} + m_d^{} P\!_L^{} \big) s \,.
\end{align}
with $F^{\eta\kappa}$ being the electromagnetic field-strength tensor, \,$P_{L,R}^{} = \big(1\mp\gamma_5^{}\big)/2$,\, and $m_{d(s)}$ the $d(s)$-quark mass.\,
The impact of new physics is parameterized according to \cite{Mertens:2011ts,Geng:2021fog,Neshatpour:2022fak,Roy:2024hqg}
\begin{align} \label{Oj}
{\cal H}_{\rm eff}^{\tt NP} \,=\, -\raisebox{2pt}{\scriptsize$\displaystyle
\sum_{\mbox{\it\scriptsize r}}$}\, C_r^{}{\cal O}_r^{} \,+\, \rm H.c. \,, & &
\end{align}
where
\begin{align} \label{npo}
{\cal O}_{7(7')}^{} & \,=\, \frac{\lambda_t^{}{\tt e}G_{\rm F}^{}m_s^{}}{4\sqrt2\,\pi^2}\,
F_{\eta\kappa}^{}\, \overline s \sigma^{\eta\kappa} P_{L(R)}^{} d \,, &
\lambda_{\tt q}^{} & \,=\, V_{{\tt q}d}^{} V_{{\tt q}s}^* \,,
\nonumber \\
{\cal O}_{9(9')}^e & \,=\, \frac{\lambda_t^{}{\tt e}^2G_{\rm F}^{}}{4\sqrt2\,\pi^2}\,
\overline s\gamma^\eta P_{L(R)}^{}d\, \overline e\gamma_\eta^{}e \,, &
{\cal O}_{S(S')}^e & \,=\, \frac{\lambda_t^{}{\tt e}^2G_{\rm F}^{}m_s^{}}{4\sqrt2\,\pi^2}\,
\overline sP_{R(L)}^{}d\, \overline ee \,,
\nonumber \\
{\cal O}_{10(10')}^e & \,=\, \frac{\lambda_t^{}{\tt e}^2G_{\rm F}^{}}{4\sqrt2\,\pi^2}\,
\overline s\gamma^\eta P_{L(R)}^{}d\, \overline e\gamma_\eta^{}\gamma_5^{}e \,, &
{\cal O}_{P(P')}^e & \,=\, \frac{\lambda_t^{}{\tt e}^2G_{\rm F}^{}m_s^{}}{4\sqrt2\,\pi^2}\,
\overline sP_{R(L)}^{}d\, \overline e\gamma_5^{}e \,.
\end{align}
The Wilson coefficients appear below as \,$C_{r\pm}^{} = C_r^{} \pm C_{r'}^{}$, \,$r = 7, 9, 10, S, P$.\,

To address the effects of $\;\!{\cal H}_{\rm eff}^{\tt SM}$ and ${\cal H}_{\rm eff}^{\tt NP}$ on
\,{\small$\mathfrak B$}$\,\to\,${\small$\mathfrak B$}$'e^+e^-$,\, we need to apply the appropriate
\,{\small$\mathfrak B$}$\,\to\,${\small$\mathfrak B$}$'$\, matrix elements of the various quark bilinears in eqs.\,\,(\ref{smsd}) and (\ref{npo}).
For \,$s\to d$\, vector and axialvector currents, we have
\begin{align} \label{VA}
\langle\mbox{\small$\mathfrak B$}{}'|\overline d\gamma^\eta s|\mbox{\small$\mathfrak B$}\rangle
& \,=\, \bar u_{\mathfrak B'}^{} \Big( \gamma^\eta f_1^{\scriptscriptstyle\mathfrak B'\mathfrak B}
+ i\sigma^{\eta\nu} f_2^{\scriptscriptstyle\mathfrak B'\mathfrak B} q_\nu^{}
+ f_3^{\scriptscriptstyle\mathfrak B'\mathfrak B} q^\eta \Big) u_{\mathfrak B}^{} \,,
\nonumber \\
\langle\mbox{\small$\mathfrak B$}{}'|\overline d\gamma^\eta\gamma_5^{}s
|\mbox{\small$\mathfrak B$}\rangle
& \,=\, \bar u_{\mathfrak B'}^{} \Big( \gamma^\eta g_1^{\scriptscriptstyle\mathfrak B'\mathfrak B}
+ i\sigma^{\eta\nu} g_2^{\scriptscriptstyle\mathfrak B'\mathfrak B} q_\nu^{}
+ g_3^{\scriptscriptstyle\mathfrak B'\mathfrak B} q^\eta \Big) \gamma_5^{} u_{\mathfrak B}^{} \,,
\end{align}
where $f_{1,2,3}^{\scriptscriptstyle\mathfrak B'\mathfrak B}$ and
$g_{1,2,3}^{\scriptscriptstyle\mathfrak B'\mathfrak B}$ are form factors, for which we adopt from $\chi$PT at leading order \cite{Tandean:2019tkm}
\begin{align}
f_1^{n\Lambda} & \,=\, -\frac{\sqrt3}{\sqrt2} \,, &
g_1^{n\Lambda} & \,=\,  \frac{-{\cal D}\!-\!3\cal F}{\sqrt6} \,, &
f_1^{\Sigma^-\Xi^-} & =\, 1 \,, ~~~ ~~ g_1^{\Sigma^-\Xi^-} =\, {\cal D}+{\cal F} \,,
\nonumber \\
f_1^{p\Sigma^+} & =\, -1 \,, & g_1^{p\Sigma^+} & =\, {\cal D}-{\cal F} \,, &
g_3^{\scriptscriptstyle\mathfrak B'\mathfrak B} & \,=\,\frac{m_{\mathfrak B}^{}+m_{\mathfrak B'}^{}}
{q^2-m_{K^0}^2} g_1^{\scriptscriptstyle\mathfrak B'\mathfrak B} \,,
\end{align}
and \,
$f_{2,3}^{\scriptscriptstyle\mathfrak B'\mathfrak B} =
g_2^{\scriptscriptstyle\mathfrak B'\mathfrak B} = 0$. \,
These lead to
\begin{align} \label{SP}
\langle\mbox{\small$\mathfrak B$}{}'|\overline d s|\mbox{\small$\mathfrak B$}\rangle & \,=\,
\frac{ f_1^{\scriptscriptstyle\mathfrak B'\mathfrak B} {\tt\,M}_-^{}\,
\bar u_{\mathfrak B'}^{} u_{\mathfrak B}^{} }{m_s^{}-m_d^{}} \,, &
\langle\mbox{\small$\mathfrak B$}{}'|\overline d\gamma_5^{}s|\mbox{\small$\mathfrak B$}\rangle
& \,=\, \frac{ g_1^{\scriptscriptstyle\mathfrak B'\mathfrak B} {\tt\,M}_+^{}\, m_{K^0\,}^2
\bar u_{\mathfrak B'}^{}\gamma_5^{}u_{\mathfrak B}^{} }{ \big(m_s^{}{+}m_d^{}\big)
\big(m_{K^0}^2\!-\!q^2\big) } \,, ~~~
\end{align}
where \,${\tt M}_\pm^{} = m_{\mathfrak B}^{}\pm m_{\mathfrak B'}^{}$.\,
For the \,$s\to d$\, tensor bilinears, we have
\begin{align} \label{<t>}
\langle\mbox{\small$\mathfrak B$}{}'|\overline d\sigma^{\eta\kappa}\big(1,\gamma_5\big)s
|\mbox{\small$\mathfrak B$}\rangle \varepsilon_\eta^*q_\kappa^{}
& \,=\, {\tt C}_\sigma^{\scriptscriptstyle\mathfrak B'\mathfrak B}\, \bar u_{\mathfrak B'}^{}
\sigma^{\eta\kappa} \big(1,\gamma_5\big)u_{\mathfrak B}^{}\,\varepsilon_\eta^*q_\kappa^{} \,, &
\end{align}
where from the quark model \cite{Gilman:1978bg,Donoghue:2022wrw}
\begin{align} \label{csigma}
{\tt C}_\sigma^{p\Sigma^+} & \,=\, c_\sigma^{} \,, &
{\tt C}_\sigma^{n\Lambda} & \,=\, \frac{-\sqrt{27}\, c_\sigma^{}}{\sqrt2} \,, &
{\tt C}_\sigma^{\Sigma^-\Xi^-} & \,=\, +5 c_\sigma^{} \,. ~~~
\end{align}
The values of the constants $\cal D$, $\cal F$, and $c_\sigma^{}$ are specified later on.

The interactions in ${\cal H}_{\rm eff}^{\tt SM}$ and ${\cal H}_{\rm eff}^{\tt NP}$ affect the amplitude for \,{\small$\mathfrak B$}$\,\to\,${\small$\mathfrak B$}$'e^+e^-$\, of the form \cite{He:2018yzu}
\begin{align} \label{MB2B'ee}
{\cal M}\, & =\, \bar u_{\mathfrak B'}^{} \big[ \big( \tilde{\textsc a} + \gamma_5^{}
\tilde{\textsc b} \big) i\sigma^{\eta\kappa} q_\kappa^{} - \gamma^\eta \big(\tilde{\textsc c}
+ \gamma_5^{}\tilde{\textsc d} \big) \big]u_{\mathfrak B}^{}\,\bar u_e^{}\gamma_\eta^{}v_{\bar e}^{}
+ \bar u_{\mathfrak B'}^{}\gamma^\eta \big(\tilde{\textsc e}+\gamma_5^{}\tilde{\textsc f}\big)
u_{\mathfrak B}^{}\, \bar u_e^{}\gamma_\eta^{}\gamma_5^{}v_{\bar e}^{}   ~~~
\nonumber \\ & ~~~ +\, \bar u_{\mathfrak B'}^{} \big( \tilde{\textsc g}+\gamma_5^{}\tilde{\textsc h}
\big) u_{\mathfrak B}^{}\,\bar u_e^{}v_{\bar e}^{} + \bar u_{\mathfrak B'}^{} \big(\tilde{\textsc j}
+ \gamma_5^{} \tilde{\textsc k} \big) u_{\mathfrak B}^{}\, \bar u_e^{}\gamma_5^{}v_{\bar e}^{} \,.
\end{align}
Explicitly, they translate into
\begin{align} \label{A...K} &
\tilde{\textsc a} \,=\, \frac{G_{\rm F}^{}}{q^2} \bigg[ 4\alpha_{\tt 0}^{}\pi\, a_{\mathfrak B}^{}
- \frac{\alpha_{\tt Z}^{}\;\! {\tt C}_\sigma^{\scriptscriptstyle\mathfrak B'\mathfrak B} m_s^{}}
{2\sqrt2\,\pi} \big(\lambda_u^* C_{7\gamma}^{} - 2\;\! \lambda_t^*C_{7+}^*\big) \bigg] \,,
\nonumber \\ &
\tilde{\textsc b} \,=\, \frac{G_{\rm F}^{}}{q^2} \bigg[ 4\alpha_{\tt 0}^{}\pi\, b_{\mathfrak B}^{}
- \frac{\alpha_{\tt Z}^{}\;\! {\tt C}_\sigma^{\scriptscriptstyle\mathfrak B'\mathfrak B} m_s^{}}
{2\sqrt2\,\pi} \big(\lambda_u^* C_{7\gamma}^{} - 2\;\! \lambda_t^*C_{7-}^*\big) \bigg] \,,
\nonumber \\ &
\tilde{\textsc c} \,=\, G_{\rm F}^{} \Bigg[ 4\alpha_{\tt 0}^{}\pi\, c_{\mathfrak B}^{}
- \Bigg( \frac{\lambda_u^*z_{7V}^{}-\lambda_t^*y_{7V}^{}}{\sqrt2} - \frac{ \alpha_{\tt Z}^{}\;\!
\widetilde{\tt c}{}_{9+}^{e*}}{\sqrt8\,\pi} \Bigg) f_1^{\scriptscriptstyle\mathfrak B'\mathfrak B}
\Bigg] \,,
\nonumber \\ &
\tilde{\textsc d} \,=\, G_{\rm F}^{} \Bigg[ 4\alpha_{\tt 0}^{}\pi\, d_{\mathfrak B}^{} + \Bigg(
\frac{\lambda_u^*z_{7V}^{}-\lambda_t^*y_{7V}^{}}{\sqrt2} - \frac{\alpha_{\tt Z}^{}\;\!
\widetilde{\tt c}{}_{9-}^{e*}}{\sqrt8\,\pi} \Bigg) g_1^{\scriptscriptstyle\mathfrak B'\mathfrak B}
\Bigg] \,,
\\
\tilde{\textsc e} =~ & \frac{-G_{\rm F}^{}}{\sqrt2} \bigg( \lambda_t^*y_{7A}^{}
+ \frac{\alpha_{\tt Z}^{}}{2\pi}\, \widetilde{\tt c}{}_{10+}^{e*} \bigg)
f_1^{\scriptscriptstyle\mathfrak B'\mathfrak B} \,,
\hspace{3em} \tilde{\textsc f} = \frac{G_{\rm F}^{}}{\sqrt2} \bigg( \lambda_t^* y_{7A}^{}
+ \frac{\alpha_{\tt Z}^{}}{2\pi}\,\widetilde{\tt c}{}_{10-}^{e*} \bigg)
g_1^{\scriptscriptstyle\mathfrak B'\mathfrak B} \,,
\nonumber \\
\tilde{\textsc g} =~ & \frac{ \alpha_{\tt Z}^{}G_{\rm F}^{}\,\widetilde{\tt c}{}_{\!S+\;\!}^{e*}
\tt M_-^{} }{2\sqrt2\,\pi} f_1^{\scriptscriptstyle\mathfrak B'\mathfrak B} \,, ~~~ ~~
\tilde{\textsc h} = \frac{ \alpha_{\tt Z}^{}G_{\rm F}^{}\, \widetilde{\tt c}{}_{\!S-}^{e*}
m_{\!K^0\;\!}^2 \tt M_+^{} }{\sqrt8\,\pi\,\big(m_{K^0}^2{-}q^2\big)}\;\!
g_1^{\scriptscriptstyle\mathfrak B'\mathfrak B} \,, ~~~ ~~
\tilde{\textsc j} = \frac{ -\alpha_{\tt Z}^{}G_{\rm F}^{}\, \widetilde{\tt c}{}_{\!P+\;\!}^{e*}
\tt M_-^{} }{2\sqrt2\,\pi} f_1^{\scriptscriptstyle\mathfrak B'\mathfrak B} \,,
\nonumber \\ \nonumber
\tilde{\textsc k} =~ & \frac{G_{\rm F}^{}}{\sqrt2}\! \bigg[ 2 \lambda_t^* y_{7\!A}^{} m_e^{}
+ \frac{\alpha_{\tt Z}^{}}{\pi}\! \Big( \widetilde{\tt c}{}_{\!10-}^{e*} m_e^{}
+ \widetilde{\tt c}{}_{\!P-}^{e*} m_{\!K^0}^2\! \Big)\! \bigg]\!
\frac{g_1^{\scriptscriptstyle\mathfrak B'\mathfrak B\;\!} \tt M_+^{}}{q^2\!-\!m_{\!K^0}^2} \,, ~~~ ~~~~
\widetilde{\tt c}{}_r^e \,\equiv\, \lambda_t^{} C_r^e \,,
\end{align}
where \,$\alpha_0^{} = {\tt e}^2(0)/(4\pi)=1/137.036$,\, $\alpha_{\tt Z}^{} = {\tt e}^2(m_Z)/(4\pi)=1/127.93$  \cite{ParticleDataGroup:2026mpi}, and $m_d^{}$ has been neglected compared to $m_s^{}$.
The expressions for \,$\tilde{\textsc a},...,\tilde{\textsc k}$\, in the \,$\Sigma^+\to pe^+e^-$\, case agree with those obtained in ref.\,\cite{Roy:2024hqg}.
The formula for the decay rate corresponding to $\cal M$ is available from ref.\,\cite{He:2018yzu}.

\section{\textit{CP} asymmetries within and beyond SM\label{cpv-sm-np}}

If the underlying interaction does not conserve $CP$ symmetry, the differential rate $d\Gamma/dq^2$ of \,$\mbox{\small$\mathfrak B$}\to\mbox{\small$\mathfrak B$}{}'e^+e^-$\, and that of the antiparticle mode, \,$\overline{\mbox{\small$\mathfrak B$}}\to\overline{\mbox{\small$\mathfrak B$}}{}'e^+e^-$,\, could be unequal.
Hence their difference is a $CP$-violating observable, which can be written as
\begin{align} \label{G'BB'ee}
\delta & \Gamma_{{\scriptscriptstyle\mathfrak{BB}}'ee}' \,\equiv\,
\frac{ d\Gamma( \mbox{\small$\mathfrak B$}\to
\mbox{\small$\mathfrak B$}'e^+e^- ) }{dq^2} - \frac{ d\Gamma \big( \raisebox{-1pt}
{\small$\overline{\mathfrak B}$} \to \raisebox{-1pt}{$\overline{\mbox{\small$\mathfrak B$}}$}{}'
e^+e^- \big) }{dq^2}
\nonumber \\ & =\, \frac{ \big(\beta^3{-}3\beta\big)\bar\lambda\raisebox{2pt}{$^{\!\frac{1}{2}}$}
\alpha_{\tt O}^{}G_{\rm F}^2 }{8\sqrt2\,\pi^2m_{\mathfrak B}^3} \bigg\{ \!\! \begin{array}[t]{l}
\displaystyle {\rm Im} \bigg[ a_{\mathfrak B}^{\;~}{\tt M_+^{}} + \frac{c_{\mathfrak B}^{}}{3}\;\!
{\tt\hat m_+^2} + c_{\mathfrak B}^{}q^2 \bigg] {\rm\,Im} \bigg[ \lambda_t^{}y_{7V}^{}
+ \frac{\alpha_{\tt Z}}{2\pi}\;\! \widetilde{\tt c}{}_{9+}^e \bigg]
f_1^{\scriptscriptstyle\mathfrak B'\mathfrak B\;\!} \tt\hat m_-^2
\\ \displaystyle {\rm\!\!+\;Im} \bigg[ b_{\mathfrak B}^{\;~}{\tt M}_-^{}
- \frac{d_{\mathfrak B}^{}}{3}\;\! {\tt\hat m_-^2} - d_{\mathfrak B}^{}q^2 \bigg] {\rm\,Im} \bigg[
\lambda_t^{} y_{7V}^{} + \frac{\alpha_{\tt Z}}{2\pi}\;\! \widetilde{\tt c}{}_{9-}^e \bigg]
g_1^{\scriptscriptstyle\mathfrak B'\mathfrak B\;\!} \tt\hat m_+^2 \bigg\} \end{array}
\nonumber \\ & ~~~~ +\, \frac{ \big(\beta^3{-}3\beta\big)
\bar\lambda\raisebox{2pt}{$^{\!\frac{1}{2}}$} \alpha_{\tt 0}^{}\alpha_{\tt Z}^{}\;\! G_{\rm F}^2\,
\mathscr C_\sigma^{\scriptscriptstyle\mathfrak B'\mathfrak B}m_s^{} }{ 8\sqrt2\,\pi^3
m_{\mathfrak B}^3 } \! \begin{array}[t]{l} \displaystyle {\rm\bigg\{Im} \bigg[ \bar\lambda\,
\frac{2 a_{\mathfrak B}^{}}{3q^2} + \big(a_{\mathfrak B}^{}+c_{\mathfrak B}^{\;~}{\tt M_+}\big)
{\tt\hat m_-^2} \bigg] {\rm~\!\!Im\;\!} {\tt\widetilde C_{7+}^{}}
\\ \displaystyle ~{\rm+~Im} \bigg[ \bar\lambda\, \frac{2 b_{\mathfrak B}^{}}{3q^2}
+ \big(b_{\mathfrak B}^{} - d_{\mathfrak B}^{\;~}{\tt M_-^{}}\big) {\tt\hat m_+^2} \bigg]
{\rm Im\;\!} {\tt\widetilde C_{7-}^{}} \bigg\} \,, \end{array} ~~~ ~~~~
\end{align}
where
\begin{align} \label{tC7+-}
\beta = \sqrt{1-\frac{4m_e^2}{q^2}} \,, & & \bar\lambda = {\tt\hat m_-^2\hat m_+^2} \,, & &
{\tt\hat m_\pm^2} = {\tt M_\pm^2} - q^2 \,, & &
{\tt\widetilde C_{7\pm}^{}} = \lambda_t^{} C_{7\pm}^{}-\frac{\lambda_u}{2}C_{7\gamma}^* \,.
\end{align}
We can then devise the differential rate asymmetry
\begin{align} \label{DeltaBee'}
\hat\Delta_{{\scriptscriptstyle\mathfrak{BB}}'ee}' \,\equiv\,
\frac{d\hat\Delta_{{\scriptscriptstyle\mathfrak{BB}}'ee}}{dq^2} \,=\,
\frac{\tau_{\mathfrak B}^{}\, \delta\Gamma_{\!\mathfrak{BB}{}'ee}'}
{2\langle{\cal B}_{{\scriptscriptstyle\mathfrak{BB}}'ee}\rangle} \,,
\end{align}
where $\tau_{\mathfrak B}^{}$ is the lifetime of {\small$\mathfrak B$} and \,$2\langle{\cal B}_{{\scriptscriptstyle\mathfrak{BB}}'ee}\rangle = \mathscr B_{{\scriptscriptstyle\mathfrak{BB}}'ee} + \overline{\mathscr B}{}_{{\scriptscriptstyle\mathfrak{BB}}'ee}$,\, with $\mathscr B_{{\scriptscriptstyle\mathfrak{BB}}'ee}$ the branching fraction of \,{\small$\mathfrak B$}$\,\to\,${\small$\mathfrak B$}${}'e^+e^-$\, and
$\overline{\mathscr B}_{{\scriptscriptstyle\mathfrak{BB}}'ee}$ that of its antiparticle counterpart,
\,$\overline{\mbox{\small$\mathfrak B$}}\to\overline{\mbox{\small$\mathfrak B$}}{}'e^+e^-$.\,
The asymmetry $\hat\Delta_{{\scriptscriptstyle\mathfrak{BB}}'ee}$ results from integrating eq.\,(\ref{DeltaBee'}) over \,$4m_e^2\le q^2\le(m_{\mathfrak B}-m_{\mathfrak B'})^2$.\,
From eq.\,(\ref{G'BB'ee}) we conclude that, among the possible new contributions listed in eq.\,(\ref{Oj}), $\hat\Delta_{{\scriptscriptstyle\mathfrak{BB}}'ee}$ probes only Im$(\lambda_tC_{7\pm})$ and Im$\big(\lambda_t^{}C_{9\pm}^e\big)$.

For the weak radiative process \,{\small$\mathfrak B$}$\,\to\,${\small$\mathfrak B$}$'\gamma$,\,
which emits an onshell photon, the amplitude and rate are, respectively,
\begin{align}
\mathscr M_{{\scriptscriptstyle\mathfrak{BB}}'\gamma}^{} & \,=\, i{\tt e} G_{\rm F\;\!}^{}
\bar u_{\mathfrak B'}^{} \big( {\tt A}_{{\scriptscriptstyle\mathfrak{BB}}'\gamma}
+ \gamma_5^{} {\tt~\!\!B}_{{\scriptscriptstyle\mathfrak{BB}}'\gamma} \big)
\sigma^{\kappa\nu} u_{\mathfrak B}^{}\, \varepsilon_\kappa^* q_\nu^{} \,, &
\nonumber \\
\Gamma\!_{{\scriptscriptstyle\mathfrak{BB}}'\gamma}^{} & \,=\, \frac{\alpha_{\tt0}^{}G_{\rm F}^2}{2 m_{\mathfrak B}^3}
 \big(m_{\mathfrak B}^2-m_{\mathfrak B'}^2\big)\raisebox{1pt}{$^3$}
\big(|{\tt A}_{{\scriptscriptstyle\mathfrak{BB}}'\gamma}|^2
+ |{\tt B}_{{\scriptscriptstyle\mathfrak{BB}}'\gamma}|^2\big) \,, \label{B2B'g-rate}
\end{align}
where ${\tt A}_{{\scriptscriptstyle\mathfrak{BB}}'\gamma}$ and
${\tt B}_{{\scriptscriptstyle\mathfrak{BB}}'\gamma}$ comprise LD and SD components, the latter containing the SM coefficient $C_{7\gamma}$ and the NP ones $C_{7\pm}$.
Explicitly
\begin{align} \label{AgBg}
{\tt A}_{{\scriptscriptstyle\mathfrak{BB}}'\gamma}^{} & \,=\, a_{\mathfrak B}^{}(0)
+ \alpha_{\tt Z}^{\tt1/2}\, \frac{{\tt C}_\sigma^{\scriptscriptstyle\mathfrak B'\mathfrak B}\,{\tt\widetilde C_{7+}^*}m_s^{}}
{4\sqrt{2\alpha_{\tt0}^{}}~ \pi^2} \,, &
{\tt B}_{{\scriptscriptstyle\mathfrak{BB}}'\gamma}^{} & \,=\, b_{\mathfrak B}^{}(0)
+ \alpha_{\tt Z}^{\tt1/2}\, \frac{{\tt C}_\sigma^{\scriptscriptstyle\mathfrak B'\mathfrak B}\,{\tt\widetilde C_{7-}^*}m_s^{}}
{4\sqrt{2\alpha_{\tt0}^{}}~ \pi^2} \,,
\end{align}
where $a_{\mathfrak B}^{}(0)$ and $b_{\mathfrak B}^{}(0)$ are the \,$q^2=0$\, values of
$a_{\mathfrak B}^{}$ and $b_{\mathfrak B}^{}$ and $\tt\widetilde C{}_{7\pm}^{}$ is  defined
in eq.\,(\ref{tC7+-}).
The antihyperon mode
\,$\overline{\mbox{\small$\mathfrak B$}}\to\overline{\mbox{\small$\mathfrak B$}}{}'\gamma$\, has
analogous formulas, with ${\tt A}_{\mathfrak B\gamma}$ and ${\tt B}_{\mathfrak B\gamma}$
replaced by
\begin{align} \label{bAgbBg}
\overline{\tt A}_{{\scriptscriptstyle\mathfrak{BB}}'\gamma}^{} &  \,=\, -a_{\mathfrak B}^{}(0)
- \alpha_{\tt Z}^{\tt1/2}\, \frac{{\tt C}_\sigma^{\scriptscriptstyle\mathfrak B'\mathfrak B}\,{\tt\widetilde C_{7+}^{}}m_s^{}}
{4\sqrt{2\alpha_{\tt0}^{}}~ \pi^2} \,, &
\overline{\tt B}_{{\scriptscriptstyle\mathfrak{BB}}'\gamma}^{} &  \,=\, b_{\mathfrak B}^{}(0)
+ \alpha_{\tt Z}^{\tt1/2}\, \frac{{\tt C}_\sigma^{\scriptscriptstyle\mathfrak B'\mathfrak B}\,{\tt\widetilde C_{7-}^{}}m_s^{}}
{4\sqrt{2\alpha_{\tt0}^{}}~ \pi^2} \,.
\end{align}
Accordingly, we have the $CP$-violating observables
\begin{align} \label{A,Delta}
\hat\Delta_{{\scriptscriptstyle\mathfrak{BB}}'\gamma}^{}\, & \equiv\, \frac{
\mathscr B_{{\scriptscriptstyle\mathfrak{BB}}'\gamma}
- \overline{\mathscr B}_{{\scriptscriptstyle\mathfrak{BB}}'\gamma} }
{ \mathscr B_{{\scriptscriptstyle\mathfrak{BB}}'\gamma}
+ \overline{\mathscr B}_{{\scriptscriptstyle\mathfrak{BB}}'\gamma} }
\,=\, \frac{ \Gamma\!_{{\scriptscriptstyle\mathfrak{BB}}'\gamma}
- \overline\Gamma\!_{{\scriptscriptstyle\mathfrak{BB}}'\gamma}}
{ \Gamma\!_{{\scriptscriptstyle\mathfrak{BB}}'\gamma}
+ \overline\Gamma_{{\scriptscriptstyle\mathfrak{BB}}'\gamma} } \,, &
\hat A_{{\scriptscriptstyle\mathfrak{BB}}'\gamma}^{} & \,\equiv\,
\frac{ \alpha_{{\scriptscriptstyle\mathfrak{BB}}'\gamma}^{}
+ \overline\alpha{}_{{\scriptscriptstyle\mathfrak{BB}}'\gamma}^{} }
{ \alpha_{{\scriptscriptstyle\mathfrak{BB}}'\gamma}^{}
- \overline\alpha{}_{{\scriptscriptstyle\mathfrak{BB}}'\gamma}^{} } \,, ~
\end{align}
where $\mathscr B_{{\scriptscriptstyle\mathfrak{BB}}'\gamma}$
$(\alpha_{{\scriptscriptstyle\mathfrak{BB}}'\gamma})$ is the branching fraction
(decay asymmetry parameter) of \,{\small$\mathfrak B$}$\,\to\,${\small$\mathfrak B$}$'\gamma$\,
while $\overline{\mathscr B}\!_{{\scriptscriptstyle\mathfrak{BB}}'\gamma}$
$\big(\overline\alpha\!_{{\scriptscriptstyle\mathfrak{BB}}'\gamma}\big)$ denotes the corresponding quantity of \,$\overline{\mbox{\small$\mathfrak B$}}\to\overline{\mbox{\small$\mathfrak B$}}{}'\gamma$,\, with
\begin{align}
\alpha_{{\scriptscriptstyle\mathfrak{BB}}'\gamma}^{} \,=\,
\frac{2\, {\rm Re}\big( {\tt A}_{{\scriptscriptstyle\mathfrak{BB}}'\gamma}^*\,
{\tt B}_{{\scriptscriptstyle\mathfrak{BB}}'\gamma}^{} \big) }
{|{\tt A}_{{\scriptscriptstyle\mathfrak{BB}}'\gamma}|^2
+ |{\tt B}_{{\scriptscriptstyle\mathfrak{BB}}'\gamma}|^2} \,, & &
\overline\alpha_{{\scriptscriptstyle\mathfrak{BB}}'\gamma}^{} \,=\, \frac{ 2\, {\rm Re}\big(
\overline{\tt A}{}_{{\scriptscriptstyle\mathfrak{BB}}'\gamma}^*\,
\overline{\tt B}_{{\scriptscriptstyle\mathfrak{BB}}'\gamma}^{} \big) }
{|\overline{\tt A}_{{\scriptscriptstyle\mathfrak{BB}}'\gamma}|^2
+ |\overline{\tt B}_{{\scriptscriptstyle\mathfrak{BB}}'\gamma}|^2} \,. & ~~~ ~~~~
\end{align}
Defining
\,$2\langle{\cal B}_{{\scriptscriptstyle\mathfrak{BB}}'\gamma}\rangle =
\mathscr B_{{\scriptscriptstyle\mathfrak{BB}}'\gamma}
+ \overline{\mathscr B}_{{\scriptscriptstyle\mathfrak{BB}}'\gamma}$\,
and
\,$2\langle\alpha_{{\scriptscriptstyle\mathfrak{BB}}'\gamma}\rangle =
\alpha_{{\scriptscriptstyle\mathfrak{BB}}'\gamma}^{}
- \overline\alpha_{{\scriptscriptstyle\mathfrak{BB}}'\gamma}^{}$,\, we expand
$\hat\Delta_{{\scriptscriptstyle\mathfrak{BB}}'\gamma}$ and
$\hat A_{{\scriptscriptstyle\mathfrak{BB}}'\gamma}$ in terms of Im$\tt\,\widetilde C_{7\pm}$ into
\begin{align} \label{Delta-g}
\hat\Delta_{{\scriptscriptstyle\mathfrak{BB}}'\gamma}^{} & \,=\, \frac{ \sqrt{\alpha_{\tt0}^{}
\alpha_{\tt Z}^{}} \big(m_{\mathfrak B}^2 \!-\! m_{\mathfrak B'}^2\big)\raisebox{1pt}{$^3$}
\tau_{\mathfrak B}^{} G_{\rm F\;\!}^2 {\tt C}_\sigma^{\scriptscriptstyle\mathfrak B'\mathfrak B}
m_s^{} }{ 4\sqrt2\, \pi^2\, \langle{\cal B}_{{\scriptscriptstyle\mathfrak{BB}}'\gamma}\rangle
\, m_{\mathfrak B}^3} \Big[ {\rm-Im}\;\!a_{\mathfrak B}^{}(0) {\rm\,Im}{\tt\;\!\widetilde C_{7+}^{}}
- {\rm Im}\;\!b_{\mathfrak B}^{}(0) {\rm\,Im}{\tt\;\!\widetilde C_{7-}^{}} \Big] \,,
\end{align}
which is exact, and
\begin{equation} \label{A-g} \!\!\!\begin{array}[b]{rl} \displaystyle
\hat A_{{\scriptscriptstyle\mathfrak{BB}}'\gamma}^{} \,=\, \frac{
-\big(m_{\mathfrak B}^2\!-\!m_{\mathfrak B'}^2\big)\raisebox{1pt}{$^3$}\, \tau_{\mathfrak B}^{}\,
G_{\rm F}^2 }{ \langle\alpha_{{\scriptscriptstyle\mathfrak{BB}}'\gamma}\rangle\,
\langle{\cal B}_{{\scriptscriptstyle\mathfrak{BB}}'\gamma}\rangle\, m_{\mathfrak B}^3 }
\Bigg\{ \!\!\! & \displaystyle \frac{ {\rm Im}\;\!a_{\mathfrak B}^{}(0)
{\rm\,Im}\;\!{\tt\widetilde C_{7-}} \!+\! {\rm\,Im}\;\!b_{\mathfrak B}^{}(0)
{\rm\,Im}\;\!\tt\widetilde C_{7+} }{4\sqrt2\,\pi^2} \sqrt{\alpha_{\tt0}^{}\alpha_{\tt Z}^{}}
{\tt~C}_\sigma^{\scriptscriptstyle\mathfrak B'\mathfrak B} m_s^{}
\\ & \!+\,\, \alpha_{\tt0}^{}\, \hat\Delta_{{\scriptscriptstyle\mathfrak{BB}}'\gamma}^{}
{\,\rm Re} \big[a_{\mathfrak B}^*(0)\, b_{\mathfrak B}^{}(0)\big] \Bigg\} \end{array}
\end{equation}
to first order in Im$(\lambda_tC_{7\pm})$.

In the numerical treatment of these asymmetries we employ \,$z_{7V}^{} = (-0.05\pm0.02)\alpha_{\tt Z}^{}$,\linebreak  $y_{7V}^{} \!=\! (0.73 \!\pm\! 0.04)\alpha_{\tt Z}^{}$,\, $y_{7A}^{} \!=\! (-0.68 \!\pm\! 0.03)\alpha_{\tt Z}^{}$  \cite{Buchalla:1995vs,Isidori:2004rb,Neshatpour:2022fak}, $C_{7\gamma} \!=\! 0.72 \!-\! 0.79i{\rm\;\!Im}\lambda_t^{}$ \cite{Roy:2024hqg,Nielsen:1995dp,Tandean:1999mg},\linebreak and \,$m_s=(123\pm1)$~MeV,\, all evaluated at the renormalization scale of 1\,\,GeV.
Moreover, we adopt the CKM parameters leading to \,$\lambda_t = [-3.24\pm0.14+(1.44\pm0.07)i] \!\times\! 10^{-4}$,\, hadron and lepton masses, and hyperon lifetimes from ref.\,\cite{ParticleDataGroup:2026mpi}, \,$c_\sigma^{}=0.30\pm0.09$\, estimated in the quark model \cite{Gilman:1978bg,Donoghue:2022wrw}, and \,${\cal D} = 0.81\pm0.01$  and \,${\cal F} = 0.47\pm0.01$\, from fitting to the data on semileptonic baryon decays \cite{ParticleDataGroup:2026mpi,Cabibbo:2003cu}.
The values of some other pertinent constants are collected in the appendixes.
The uncertainties of the various input parameters and quoted data are propagated to the branching-fraction and $CP$-asymmetry outcomes presented below.

Also germane are data on the branching fractions of the modes of interest and on the decay asymmetry parameter $\alpha$ of the weak radiative ones.
For the latter, we have \cite{ParticleDataGroup:2026mpi,BESIII:2022rgl,BGHLQR:1986kqv,E761:1993unn}
\begin{align} \label{BL2ng-data}
\mathscr B_{\Lambda n\gamma}^{\tt exp} \,=\,~ & (8.20\pm0.45\pm0.66)\!\times\!10^{-4} , &
\overline{\mathscr B}{}_{\Lambda n\gamma}^{\tt exp} \,=\,~ &
(8.62\pm 0.71\pm 0.84)\!\times\!10^{-4} ,
\\ \label{aL2ng-data}
\alpha_{\Lambda n\gamma}^{\tt exp} \,=\,~ & {-}0.13\pm 0.13\pm 0.03 \,, &
\overline\alpha{}_{\Lambda n\gamma}^{\tt exp} \,=\,~ & 0.21\pm 0.15\pm 0.06 \,,
\\ \label{X2Sg-data}
\mathscr B_{\Xi^-\Sigma^-\gamma}^{\tt exp} & \,=\, (1.27\pm0.23)\!\times\!10^{-4} \,, &
\alpha_{\Xi^-\Sigma^-\gamma}^{\tt exp} & \,=\, 1.0\pm 1.3 \,,
\end{align}
but the experimental information on \,$\overline\Xi{}^+\to\overline\Sigma{}^+\gamma$ is still lacking.
The BESIII findings~\cite{BESIII:2022rgl} cited in eqs.\,\,(\ref{BL2ng-data}) and (\ref{aL2ng-data}) can be translated into
\begin{align} \label{DL2ng-data}
\hat\Delta_{\Lambda n\gamma}^{\tt exp} & \,=\, -0.025 \pm 0.080 \,, &
\hat A_{\Lambda n\gamma}^{\tt exp} & \,=\, -0.24 \pm 0.61 \,, &
\end{align}
where the statistical and systematic errors have been summed in quadrature and the systematics of $\Lambda$ and its antiparticle taken to be uncorrelated.
Despite being consistent with zero, $\hat\Delta_{\Lambda n\gamma}^{\tt exp}$ can already bring about improved limitation on Im$(\lambda_tC_{7-})$, as will be shown shortly.

The dielectron channels, \,$\Lambda\to ne^+e^-$\, and \,$\Xi^-\to\Sigma^-e^+e^-$,\, have not yet been observed.
Consequently, in the following we normalize their rate asymmetries with their branching fractions from the SM, which can be determined after combining the SD contributions proceeding from eq.\,(\ref{smsd}) and the LD ones, including the form factors calculated in appendices \ref{L2ng-ff} and~\ref{X2Sg-ff}.
The results are
\begin{align} \label{BLnee-sm}
\mathscr B_{\Lambda nee}^{\tt SM} & \,=\, (6.1\pm0.5)\!\times\!10^{-6} \,=\, \overline{\mathscr B}{}_{\Lambda nee}^{\tt SM} + (1.1\pm0.3)\!\times\!10^{-10} \,,
\\ \label{BXSee-sm}
\mathscr B_{\Xi^-\Sigma^-ee}^{\tt SM} & \,=\, (9.0\pm0.3)\!\times\!10^{-7} \,=\, \overline{\mathscr B}{}_{\Xi^-\Sigma^-ee}^{\tt SM} + (2.4\pm0.8)\!\times\!10^{-11} \,, ~~~ ~~
\end{align}
where the errors reflect the uncertainties of the input parameters.
If \,$c=d=0$,\, we would obtain instead \,$\mathscr B_{\Lambda nee}^{{\tt SM,}\,c\,=\,d\,=\,0} = (6.4\pm0.5)\!\times\!10^{-6}$\, and \,$\mathscr B_{\Xi^-\Sigma^-ee}^{{\tt SM,}\,c\,=\,d\,=\,0} = (9.2\pm0.3)\!\times\!10^{-7}$.\,
The SD parts alone are far smaller: $\mathscr B_{\Lambda nee}^{\tt SM,SD}=(3.9\pm2.4)\!\times\!10^{-9}$\, and \,$\mathscr B_{\Xi^-\Sigma^-ee}^{\tt SM,SD}=(1.6\pm1.0)\!\times\!10^{-9}$.\,
We can additionally mention here that the SM branching-fractions of the weak radiative modes solely due to the SD contributions, \,$\mathscr B_{\Lambda n\gamma}^{\tt SM,SD}=(5.3\pm3.2)\!\times\!10^{-7}$\, and \,$\mathscr B_{\Xi^-\Sigma^-\gamma}^{\tt SM,SD}=(2.4\pm1.4)\!\times\!10^{-7}$, computed using eqs.\,\,(\ref{B2B'g-rate}) and (\ref{AgBg}), are also suppressed in view of the measurements in eqs.\,\,(\ref{BL2ng-data}) and (\ref{X2Sg-data}).

Putting things together, from eq.\,(\ref{DeltaBee'}) we arrive at the rate asymmetry of \,$\Lambda\to ne^+e^-$,\, expressed in terms of Im$(\lambda_tC_{7\pm})$ and Im$\big(\lambda_t^{}C_{9\pm}^e\big)$,
\begin{align} \label{DeltaLnee'}
\hat\Delta_{\Lambda nee}^{}\, & =\, \frac{\mathscr B_{\Lambda nee} - \overline{\mathscr B}{}_{\Lambda nee}}
{\mathscr B_{\Lambda nee}^{\tt SM} + \overline{\mathscr B}{}_{\Lambda nee}^{\tt SM}}
\nonumber \\ & =\, \hat\Delta_{\Lambda nee}^{\tt SM} - (0.041\pm0.013) {\rm\,Im} (\lambda_tC_{7+})
+ (0.41\pm0.13) {\rm\,Im} (\lambda_tC_{7-})
\nonumber \\ & ~~~ -\, (0.0014\pm0.0001) {\rm\,Im} \big(\lambda_t^{}C_{9+}^e\big)
+ (0.022\pm0.002) {\rm\,Im} \big(\lambda_t^{}C_{9-}^e\big) \,, ~~~ ~~~~
\\ \raisebox{3ex}{} \label{smDeltaLee}
\hat\Delta_{\Lambda nee}^{\tt SM} & \,=\, (9.3 \pm 2.5)\!\times\! 10^{-6} \,,
\end{align}
having set \,$2\langle{\cal B}_{\Lambda nee}\rangle$\, in the denominator of $\hat\Delta_{\Lambda nee}$ to \,$\mathscr B_{\Lambda nee}^{\tt SM} + \overline{\mathscr B}{}_{\Lambda nee}^{\tt SM}$\, from eq.\,(\ref{BLnee-sm}).
Likewise, from eqs.\,\,(\ref{Delta-g}) and (\ref{A-g}) we derive for \,$\Lambda\to n\gamma$\,
\begin{align} \label{DeltaLng'}
\hat\Delta_{\Lambda n\gamma}^{}\, & =\, \hat\Delta_{\Lambda n\gamma}^{\tt SM}
- (0.042\pm0.013) {\rm\,Im} (\lambda_tC_{7+}) + (0.38\pm0.12) {\rm\,Im} (\lambda_tC_{7-}) \,, ~~~ ~~
\\ \label{ALng'} \raisebox{3ex}{}
\langle\alpha_{\Lambda n\gamma}\rangle \hat A_{\Lambda n\gamma}^{}\, & =\, \langle\alpha_{\Lambda n\gamma}\rangle \hat A_{\Lambda n\gamma}^{\tt SM} + (0.38\pm0.12)
{\rm\,Im}(\lambda_tC_{7+}) + (0.020\pm0.039){\rm\,Im}(\lambda_tC_{7-}) \,,
\end{align}
where
\begin{align} \label{smDeltaLng}
\hat\Delta_{\Lambda n\gamma}^{\tt SM}\, & =\, (-4.3 \pm 1.3)\!\times\! 10^{-6} \,, \hspace{3em} \langle\alpha_{\Lambda n\gamma}\rangle \hat A_{\Lambda n\gamma}^{\tt SM} \,=\, (-5.0 \pm 1.6)\!\times\! 10^{-6} \,. ~
\end{align}
In $\hat\Delta_{\Lambda n\gamma}$ and $\hat A_{\Lambda n\gamma}$ we have chosen \,$\langle{\cal B}_{\Lambda n\gamma}\rangle = \big\langle{\cal B}_{\Lambda n\gamma}^{\tt exp}\big\rangle = (8.35\pm0.65)\!\times\! 10^{-4}$, from eq.\,(\ref{BL2ng-data}), but in $\hat A_{\Lambda n\gamma}$ we have factored out \,$\langle\alpha_{\Lambda n\gamma}\rangle$ because \,$\big\langle\alpha_{\Lambda n\gamma}^{\tt exp}\big\rangle = -0.16\pm0.10$,\, from eq.\,(\ref{aL2ng-data}), has a big relative error.
Implementing analogous steps in the $\Xi^-$ case, we get
\begin{align} \label{DeltaXee'}
\hat\Delta_{\Xi^-\Sigma^-ee}^{}\, & =\, \frac{\mathscr B_{\Xi^-\Sigma^-ee} - \overline{\mathscr B}{}_{\Xi^-\Sigma^-ee}}{\mathscr B_{\Xi^-\Sigma^-ee}^{\tt SM} + \overline{\mathscr B}{}_{\Xi^-\Sigma^-ee}^{\tt SM}}
\nonumber \\ & =\, \hat\Delta_{\Xi^-\Sigma^-ee}^{\tt SM} + (0.048\pm 0.014) {\rm\,Im} (\lambda_tC_{7+})
+ (0.79\pm 0.24) {\rm\,Im} (\lambda_tC_{7-})
\nonumber \\ & ~~~ +~ (0.00076\pm 0.00001) {\rm\,Im} \big(\lambda_t^{}C_{9+}^e\big) + (0.035\pm 0.001) {\rm\,Im} \big(\lambda_t^{}C_{9-}^e\big) \,,
\\ \label{DeltaXSg'} \raisebox{4ex}{}
\hat\Delta_{\Xi^-\Sigma^-\gamma}^{}\, & =\, \hat\Delta_{\Xi^-\Sigma^-\gamma}^{\tt SM} + (0.052\pm 0.018) {\rm\,Im} (\lambda_tC_{7+}) + (0.78 \pm 0.28) {\rm\,Im} (\lambda_tC_{7-}) \,,
\\ \label{AXSg'} \raisebox{4ex}{}
\langle\alpha_{\Xi^-\Sigma^-\gamma}\rangle & \hat A_{\Xi^-\Sigma^-\gamma}^{} \,=\, \langle\alpha_{\Xi^-\Sigma^-\gamma}\rangle \hat A_{\Xi^-\Sigma^-\gamma}^{\tt SM} + (0.78 \pm 0.28) {\rm\,Im} (\lambda_tC_{7+})
\nonumber \\ & \hspace{4em} ~\,-\, (0.079 \pm 0.045) {\rm\,Im} (\lambda_tC_{7-}) \,,
\end{align}
where
\begin{align}
\hat\Delta_{\Xi^-\Sigma^-ee}^{\tt SM}\, & =\, (1.3 \pm 0.4)\!\times\! 10^{-5} \,, ~~~ ~~~~
\hat\Delta_{\Xi^-\Sigma^-\gamma}^{\tt SM} \,=\, (-1.0 \pm 0.4)\!\times\! 10^{-5} \,, ~~~ ~~~~
\\ \label{smAXSg}
\langle\alpha_{\Xi^-\Sigma^-\gamma}\rangle & \hat A_{\Xi^-\Sigma^-\gamma}^{\tt SM} \,=\, (-8.7 \pm 2.9)\!\times\! 10^{-6} \,.
\end{align}
and in $\hat\Delta_{\Xi^-\Sigma^-\gamma}$ and $\hat A_{\Xi^-\Sigma^-\gamma}$ we have approximated
$\big\langle{\cal B}_{\Xi^-\Sigma^-\gamma}\big\rangle$ with $\mathscr B_{\Xi^-\Sigma^-\gamma}^{\tt exp}$\, from eq.\,(\ref{X2Sg-data}).

In the previous paragraph we notice that the Im\,$(\lambda_tC_{7-})$ $\big($Im\,$(\lambda_tC_{7+})\big)$ portions of the different $\hat\Delta$ $(\hat A)$ equations have numerical prefactors that are roughly 10 or more times those of the other coefficients.
This can be understood from eqs.\,\,(\ref{G'BB'ee}), (\ref{Delta-g}), and (\ref{A-g}) which reveal that Im\,$b_{\mathfrak B}$ occurs in all of the enhanced terms.
The prominent role played by Im\,$b_{\Lambda,\Xi}^{}$ has been recognized in section \ref{ld+sd} and illustrated in tables \ref{imaimb} and \ref{BF}, along with figures \ref{L2ng:im-abcd} and~\ref{Xm2Smg:im-abcd}.
In~$\hat\Delta_{{\scriptscriptstyle\mathfrak{BB}}'ee}$, another reason for the amplification is the factor $1/q^2$ which is absent from the Im\,$\big(\lambda_t^{}C_{9\pm}^e\big)$ parts.
In the comparison with \,$\Sigma^+\to pe^+e^-,p\gamma$,\, the enlargement of the $\Lambda$ and $\Xi^-$ asymmetries is attributable mainly to the greater absorptive phases of $b_{\Lambda,\Xi}$ and also to the constant
${\tt C}_\sigma^{\scriptscriptstyle\mathfrak B'\mathfrak B}$ associated with the matrix elements in eq.\,(\ref{<t>}) of the tensor bilinears which for $\Lambda$ and $\Xi^-$ is, respectively, 3.7 and 5 times that for $\Sigma^+$ according to eq.\,(\ref{csigma}).

It is clear that the values of Im$(\lambda_tC_{7\pm,9\pm})$ which are permitted by the relevant constraints bear on how large the asymmetries in eqs.\,\,(\ref{DeltaLnee'}), (\ref{DeltaLng'}), (\ref{ALng'}), and (\ref{DeltaXee'})-(\ref{AXSg'}) might be.
Earlier~\cite{He:2026chn}, treating these Wilson coefficients model-independently, one at a time, we reported the allowed ranges \,$-0.032<{\rm Im}(\lambda_tC_{7+})<0.017$, \,$-0.19<{\rm Im}(\lambda_tC_{7-})<0.41$,\, and \,$-0.0084<{\rm Im}\big(\lambda_t^{}C_{9+}^e\big)<0.0046$\, from kaon data \cite{Mertens:2011ts,KTEV:2000ngj,KTeV:2003sls} and \,$-77<{\rm Im}\big(\lambda_t^{}C_{9-}^e\big)<68$\, from the available \,$\Sigma^+\to pe^+e^-$\, measurement \cite{Ang:1969hg}.
As it turns out, $\hat\Delta_{\Lambda n\gamma}^{\tt exp}$ in eq.\,(\ref{DL2ng-data}) from BESIII findings \cite{BESIII:2022rgl} can already strengthen the preceding Im$(\lambda_tC_{7-})$ limit, whereas our examination of \,$\Lambda\to ne^+e^-$\, and \,$\Xi^-\to\Sigma^-e^+e^-$\, suggests that they could provide significantly better restraints on Im$\big(\lambda_t^{}C_{9-}^e\big)$.
We address this in the next paragraph.

To tighten the restriction on Im$(\lambda_tC_{7-})$, we impose \,$-0.16<\hat\Delta_{\Lambda n\gamma}<0.11$,\, which corresponds to the 90\%-CL range of $\hat\Delta_{\Lambda n\gamma}^{\tt exp}$, and drop Im$(\lambda_tC_{7+})$ from eq.\,(\ref{DeltaLng'}).
This is satisfied~by
\begin{equation} \label{c7-limit}
-0.27 \,<\, {\rm\,Im}(\lambda_tC_{7-}) \,<\, 0.18 \,. ~~~ ~~~~
\end{equation}
As for Im$\big(\lambda_t^{}C_{9-}^e\big)$, with it instead being the only NP coefficient present, we have the theoretical branching fractions
\begin{align} \label{Bth}
\mathscr B_{\Lambda nee}^{} & \,=\, \mathscr B_{\Lambda nee}^{\tt SM} + 10^{-9} \big[ (137\pm4) {\rm\,Im}\big(\lambda_t^{}C_{9-}^e\big) + (7.3\pm0.2) {\rm\,Im}^2\big(\lambda_t^{}C_{9-}^e\big) \big] \,,
\nonumber \\
\overline{\mathscr B}{}_{\Lambda nee}^{} & \,=\, \mathscr B_{\Lambda nee}^{\tt SM} + 10^{-9} \big[ (-137\pm4) {\rm\,Im}\big(\lambda_t^{}C_{9-}^e\big) + (7.3\pm0.2) {\rm\,Im}^2\big(\lambda_t^{}C_{9-}^e\big) \big] \,,
\nonumber \\
\mathscr B_{\Xi^-\Sigma^-ee}^{} & \,=\, \mathscr B_{\Xi^-\Sigma^-ee}^{\tt SM} + 10^{-9} \big[ (31.9\pm0.8) {\rm\,Im}\big(\lambda_t^{}C_{9-}^e\big) + (1.77\pm0.04) {\rm\,Im}^2\big(\lambda_t^{}C_{9-}^e\big) \big] \,,
\nonumber \\
\overline{\mathscr B}{}_{\Xi^-\Sigma^-ee}^{} & \,=\, \mathscr B_{\Xi^-\Sigma^-ee}^{\tt SM} + 10^{-9} \big[ (-31.9\pm0.8) {\rm\,Im}\big(\lambda_t^{}C_{9-}^e\big) + (1.77\pm0.04) {\rm\,Im}^2\big(\lambda_t^{}C_{9-}^e\big) \big] \,.
\end{align}
Since empirical information on these modes is still nonexistent, and since they are anticipated to be greatly dominated by their LD components, we can assume as a benchmark that in the future $\mathscr B_{\Lambda nee}$, $\overline{\mathscr B}{}_{\Lambda nee}$, $\mathscr B_{\Xi^-\Sigma^-ee}$, and $\overline{\mathscr B}{}_{\Xi^-\Sigma^-ee}$ will be measured to be their SM values with errors of 10\%.
All this translates into
\begin{equation} \label{c9-limit}
-4.0 \,<\, {\rm\,Im}\big(\lambda_t^{}C_{9-}^e\big) \,<\, 4.0 \,, ~~~ ~~~~
\end{equation}
which is of similar order to the limit deduced from high-$p_T^{}$ LHC measurements \cite{Roy:2024avj}.
Upon comparing eqs.\,\,(\ref{c7-limit}) and (\ref{c9-limit}) to the aforementioned bounds from ref.\,\cite{He:2026chn}, we update the allowed ranges of the different coefficients in table \ref{wcconst}.

\begin{table}[b!]\bigskip\centering\setlength{\tabcolsep}{2ex}
\begin{tabular}{|c|c|}\hline WC & 90\%-CL range \\ \hline
Im$(\lambda_tC_{7+})\vphantom{|_|^{|_|}}$ & $(-0.032,\, 0.017)$ \\
Im$(\lambda_tC_{7-})\vphantom{|_|^|}$ & $(-0.19,\, 0.18)$ \\
Im$\big(\lambda_t^{}C_{9+}^e\big)\vphantom{|_|^|}$       & $(-0.0084,\, 0.0046)$  \\
Im$\big(\lambda_t^{}C_{9-}^e\big)\vphantom{|_{|^|}^|}$ & $(-4.0,\, 4.0)$ \\ \hline
\end{tabular}
\caption{The allowed [90\% confidence level (CL)] ranges of the imaginary parts of the Wilson
coefficients (WCs) extracted model-independently if merely one of them is contributing at a time.
The $C_{7+}^{}$ and $C_{9+}^e$ entries have been inferred from \,$K_L{\to\;\!}\pi^0e^+e^-$\, measurements \cite{He:2026chn,Mertens:2011ts,KTEV:2000ngj,KTeV:2003sls} and the $C_{7-}$ one from \,$K^+{\to\;\!}\pi^+\pi^0\gamma$ \cite{He:2026chn,Mertens:2011ts} and \,$\Lambda\to n\gamma$ \cite{BESIII:2022rgl} data.
The $C_{9-}^e$ entry follows from assuming future rate measurements of the dielectron channels within 10\% of the SM expectations.}\label{wcconst}
\end{table}

\begin{table}[b!]\bigskip\centering\setlength{\tabcolsep}{0pt}\begin{tabular}{|c|ccc|} \hline \footnotesize~Source~ & $\hat\Delta_{\Lambda nee}^{}\vphantom{|_{|_o^|}^{|_|^o}}$ & $\hat\Delta_{\Lambda n\gamma}^{}$ & $\langle\alpha_{\Lambda n\gamma}\rangle\;\! \hat A_{\Lambda n\gamma}^{}$ \\ \hline SM &
$~~(9.3 \pm 2.5)\!\times\! 10^{-6}~~$ & ~~~$(-4.3 \pm 1.3)\!\times\! 10^{-6}$~~~ &
$(-5.0 \pm 1.6)\!\times\! 10^{-6}
\vphantom{\sum_|^{|^|}}$\\
$C_{7+}$ & $(-1.0,\;\! 2.0) \!\times\! 10^{-3}$ & $(-1.1,\;\! 2.0) \!\times\! 10^{-3}$ & $~(-0.018,\;\! 0.0097)\vphantom{\sum^|}~$\\
$C_{7-}$ & $(-0.12,\;\! 0.11)$ & $(-0.11,\;\! 0.10)$ & $(-0.016,\;\! 0.015)\vphantom{\sum^|}$ \\
$C_{9+}$ & $(-0.23,\;\! 2.7) \!\times\! 10^{-5}\vphantom{\sum^|}$ & $-$ & $-$ \\
$C_{9-}$ & $(-0.10,\;\! 0.10)\vphantom{\sum_|^|}$ & $-$ & $-$ \\ \hline\hline

\footnotesize~Source~ & $\hat\Delta_{\Xi^-\Sigma^-ee}\vphantom{|_{|_o^|}^{|_|^o}}$ & $\hat\Delta_{\Xi^-\Sigma^-\gamma}$
& $\langle\alpha_{\Xi^-\Sigma^-\gamma}\rangle\;\! \hat A_{\Xi^-\Sigma^-\gamma}$ \\ \hline SM &
$~(1.3 \pm 0.4)\!\times\! 10^{-5}~$ & $~~(-1.0\pm 0.4)\!\times\! 10^{-5}~~$ &
$~(-8.7 \pm 2.9)\!\times\! 10^{-6}\vphantom{\sum_|^{|^|}}~$\\
$C_{7+}$ & $(-2.3,\;\! 1.2) \!\times\! 10^{-3}$ & $(-2.6,\;\! 1.4) \!\times\! 10^{-3}$ &
$(-0.039,\;\! 0.021)\vphantom{\sum^|}$\\
$C_{7-}$ & $(-0.22,\;\! 0.21)$ & $(-0.23,\;\! 0.22)$ & $(-0.027,\;\! 0.029)\vphantom{\sum^|}$\\
$C_{9+}$ & ~ $(-0.01,\;\! 2.4) \!\times\! 10^{-5}\vphantom{\sum^|}$ ~ & $-$ & $-$ \\
$C_{9-}$ & $(-0.15,\;\! 0.15)\vphantom{\sum_|^|}$ & $-$ & $-$ \\ \hline\hline

\footnotesize~Source~ & $\hat\Delta_{\Sigma^+pee}\vphantom{|_|^{|^|}}$ & $\hat\Delta_{\Sigma^+p\gamma}$
& $\hat A_{\Sigma^+p\gamma}$ \\ \hline
SM & $~(4.5 \pm 0.4)\!\times\! 10^{-6}~$ & $~~(1.1\pm 0.3)\!\times\! 10^{-7}~~$ & ~ $(-2.8 \pm 0.9)\!\times\! 10^{-7}
\vphantom{\sum_|^{|^|}}$ ~ \\
$C_{7+}$ & $(-0.60,\;\! 1.1)\!\times\!10^{-3}$ & $(-0.56,\;\! 1.1)\!\times\!10^{-3}$ & $(-1.9,\;\! 1.0)
\!\times\! 10^{-4} \vphantom{\sum^|}$\\
$C_{7-}$ & $(-4.2,\;\! 4.0)\!\times\!10^{-3}$ & $(-3.7,\;\! 3.5)\!\times\!10^{-3}$ & $(-5.5,\;\! 5.2)\!\times\!10^{-3} \vphantom{\sum^|}$\\
$C_{9+}$ & $(-4.1,\;\! 3.0) \!\times\! 10^{-5}$ & $-$ &
$-\vphantom{\sum^|}$\\
$C_{9-}$ & $(-7.5,\;\! 7.5)\!\times\!10^{-3}$ & $-$ & $-\vphantom{\sum_|^|}$\\ \hline
\end{tabular}\caption{The values of the $CP$-violating observables in \,$\Lambda{\,\to\,}ne^+e^-,n\gamma$,
\,$\Xi^-{\,\to\,}\Sigma^-e^+e^-,\Sigma^-\gamma$, and
\,$\Sigma^+{\,\to\,}pe^+e^-,p\gamma$\, $\big(\hat\Delta$ and $\hat A\big)$ that can be induced by the SM and by the SM in conjunction with possible new-physics encoded in the different Wilson coefficients $C_{7\pm,9\pm}$, considered model-independently and one at a~time, within their respective intervals listed in table \ref{wcconst}.} \label{cpa-sm-np}
\end{table}

\begin{figure}[h]\bigskip\centering \includegraphics[width=0.95\textwidth]{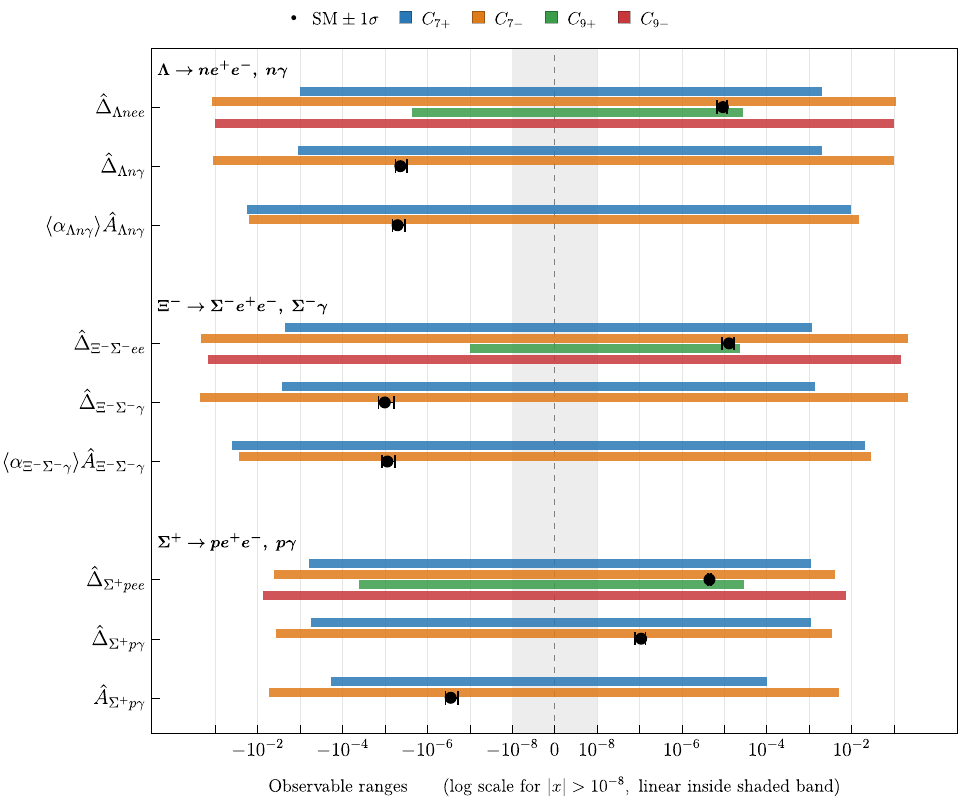}\vspace{-1ex}\caption{Pictorial ranges of the $CP$ asymmetries quoted in table \ref{cpa-sm-np}, indicating the immense potential room for new physics that can still be probed by these observables.} \label{fig:cpasym}
\end{figure}

Applying the results in table \ref{wcconst} to eqs.\,\,(\ref{DeltaLnee'}), (\ref{DeltaLng'}), (\ref{ALng'}), and (\ref{DeltaXee'})-(\ref{AXSg'}), taking into account the 90\%-CL ranges of the numbers in these formulas, and supposing as before that only one of Im$(\lambda_tC_{7\pm,9\pm})$ is nonzero at a time, we arrive at the predictions gathered in table\,\,\ref{cpa-sm-np}.
For comparison and completeness, we have included the $\Sigma^+$ results, revised from those of ref.\,\cite{He:2026chn}.
In figure \ref{fig:cpasym} we show a pictorial version of table\,\,\ref{cpa-sm-np}.

\section{Conclusions\label{concl}}

We have explored how large $CP$ violation might be in the hyperon decays \,$\Lambda\to ne^+e^-,n\gamma$\, and \,$\Xi^-\to\Sigma^-e^+e^-,\Sigma^-\gamma$\, within and beyond the SM.
The imaginary parts of their long-distance amplitudes, which we have computed from unitarity using chiral perturbation theory, play an essential role by supplying large strong phases that can help augment the $CP$-odd observables of interest.
With these imaginary parts, plus some reasonable assumptions about the real parts, we additionally predict the SM branching-fractions of the dielectron modes to be
\,$\mathscr B_{\Lambda nee}^{\tt SM} = (6.1\pm0.5)\!\times\!10^{-6}$\, and \,$
\mathscr B_{\Xi^-\Sigma^-ee}^{\tt SM} = (9.0\pm0.3)\!\times\!10^{-7}$.\,

We treat the possible quark operators due to new physics model-independently and one at a time.
With this approach, we learn that the dipole-operator Wilson coefficient Im$(\lambda_tC_{7-})$, as currently constrained by kaon data and the BESIII measurement of \,$\Lambda\to n\gamma$,\, can raise the rate asymmetries $(\hat\Delta)$ to roughly 10\% and 20\% in the $\Lambda$ and $\Xi^-$ modes, respectively, and the $\alpha$-based asymmetries $(\alpha\hat A)$ to {\footnotesize$\sim\,$}2\% and {\footnotesize$\sim\,$}3\%.
The other dipole coefficient, Im$(\lambda_tC_{7+})$, can boost $|\alpha\hat A|$ to {\footnotesize$\sim\,$}2\% and {\footnotesize$\sim\,$}4\% in \,$\Lambda\to n\gamma$\, and \,$\Xi^-\to\Sigma^-\gamma$,\, respectively, but increase $|\hat\Delta|$ merely to under 0.003.
The dielectron modes via $\hat\Delta$ are also sensitive to the semileptonic coefficient Im$\big(\lambda_tC_{9-}^e\big)$, at a slightly lesser degree than Im$(\lambda_tC_{7-})$.
The numbers owing to Im$(\lambda_tC_{7-})$ far exceed the corresponding SM expectations, which are at most a few times $10^{-5}$ in size, and are bigger than their $\Sigma^+$ counterparts by 30 to 50 times.
These findings have exposed the substantial space for new physics that remains to be inspected with $\hat\Delta$ and $\hat A$ in these processes, which hopefully serves to motivate further the experimental pursuits of them.

The BESIII data on \,$\Lambda\to n\gamma$\, has already improved the upper limit on  Im$(\lambda_tC_{7-})$.
Needless to say, more precise measurements of this channel, a better determination of the decay parameter $\alpha_{\Xi^-\Sigma^-\gamma}$, and a first measurement of the antihyperon mode \,$\overline\Xi{}^+\to\overline\Sigma{}^+\gamma$\, are highly desirable.
Observations of \,$\Lambda\to ne^+e^-$\, and \,$\Xi^-\to\Sigma^-e^+e^-$\, would test the predicted rates and offer opportunities to investigate their $CP$ asymmetries.
These may be targets for future endeavors such as PANDA \cite{PANDA:2021ozp} and the proposed Super Tau Charm Facility \cite{Achasov:2023gey}.

\acknowledgments

We thank Xiaorong Zhou for useful communications.
J.T. thanks the Tsung-Dao Lee Institute, Shanghai Jiao Tong University, for kind hospitality and support during the completion of this research.
X.-G.H. was supported by the Fundamental Research Funds for the Central Universities, by the National Natural Science Foundation of the People's Republic of China (Nos. 12090064, 12375088, and W2441004).

\appendix

\section{Form factors in long-distance amplitude for \boldmath$\Lambda\to n\gamma^*$\label{L2ng-ff}}

Their imaginary parts can be calculated reliably using unitarity from the amplitudes for the weak
nonleptonic decay \,$\Lambda\to p\pi^-$\, and for the pion photoproduction \,$p\pi^-\to n\gamma$,\,
all of which involve well-determined parameters from data.
Thus, we examine the relation
\begin{equation} \label{unitarity}
{\rm-2\,Im\,\cal M}_{\Lambda\to n\gamma^*}^{} \,=\, \raisebox{-6pt}{\large$\stackrel{\sum}
{\mbox{\scriptsize$p$\tt\,spin}}$}\,  \mbox{\Large$\int$}\! \frac{d{\tt^3P}\!_\pi^{}}
{{\tt2(2\pi)^3 E}{}_\pi^{}} \frac{d{\tt^3P}\!_p}{{\tt2(2\pi)^3 E}{}_p} (2\pi)^4
\delta^4\big({\tt P}\!_\pi^{}{\tt+P}\!_p{\tt-P}\!_\Lambda^{}\big)
{\cal M}_{p\pi\!^-\!\to n\gamma^*}^{}\;\! {\cal M}_{\Lambda\to p\pi\!^-}^{} \,,
\end{equation}
where the amplitudes inside the integral are given by
\begin{align} \raisebox{5ex}{}
{-i\cal M}_{p\pi^-\to n\gamma^*}\, & = ~ \raisebox{-7pt}{\includegraphics{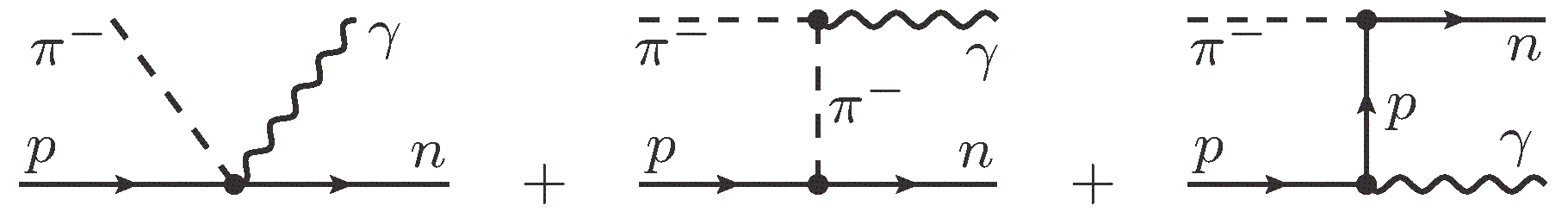}}
\nonumber \\ \raisebox{5ex}{} & =\,
\frac{\sqrt2\,({\cal D{+}F}){\tt e}\;\!m_N^{}}{f}\, \bar u_n^{} \Bigg[ \frac{ \varepsilon^*
{\cdot}\big(q-{\tt2~\!\!P}\!_\pi^{}\big) }{ q{\tt^2-2~\!\!P}\!_\pi^{}\!\!\cdot\!q }
+ \frac{m_n^{}\!-\!m_p\!-\!\slashed{\tt P}\!_\pi^{}}{q{\tt^2-2~\!\!P}\!_p\!\cdot\!q }\,
\slashed\varepsilon{}^* \Bigg] \gamma_5^{}u_p^{} \,,
\end{align}
with \,$m_N^{}=\big(m_n^{}+m_p\big)/2$,\, and \,${\cal M}_{\Lambda\to p\pi^-} = i G_{\rm F}^{}m_{\pi^+}^2 \bar u_p^{} \big({\mathbb A}_{\Lambda p\pi^-} - \gamma_5^{} {\mathbb B}_{\Lambda p\pi^-} \big) u_\Lambda^{}$.\,
The thick dots in these diagrams represent couplings furnished by the lowest-order strong chiral Lagrangian~\cite{Bijnens:1985kj}
\begin{equation} \label{Lstrong}
\begin{array}[b]{rl} {\cal L}_{\rm s}^{} \,\supset\, {\rm Tr} \!\!\! & \Big( \,\overline{\!B}\, i\gamma_\mu^{} ( \partial^\mu B + [{\cal V}^\mu, B] ) + {\cal D}\; \overline{\!B} \gamma_\mu^{} \gamma_5^{} \{{\cal A}^\mu, B\} + {\cal F}\; \overline{\!B} \gamma_\mu^{} \gamma_5^{} [{\cal A}^\mu, B]
\\ & \,+\; \tfrac{1}{4}f_{\!\pi}^2\, {\mathbb D}^\mu\hat\Sigma^\dagger \,{\mathbb D}_\mu^{} \hat\Sigma \Big) \,, \end{array} ~~~ ~~~~
\end{equation}
where the hadron and photon ($A$) fields are collected into
\begin{align} \label{fields}
B & \,=\, \frac{1}{\mbox{\small$\sqrt6$}}\! \left(\!\begin{array}{ccc} \Lambda\!+\!\mbox{\small$\sqrt3$}\,\Sigma^0 & \mbox{\small$\sqrt6$}^{\vphantom{|}}\,\Sigma^+ & \mbox{\small$\sqrt6$}\, p \\ \mbox{\small$\sqrt6$}\, \Sigma^- & \!\!\Lambda\!-\!\mbox{\small$\sqrt3$}\,\Sigma^0 & \mbox{\small$\sqrt6$}\, n \\ \mbox{\small$\sqrt6$}\, \Xi^- & \mbox{\small$\sqrt6$}\, \Xi^0 & -2\Lambda \end{array}\right) , ~~~
\varphi \,=\, \frac{1}{\mbox{\small$\sqrt3$}}\! \left(\!\begin{array}{ccc} \eta_8^{}\!+\!\mbox{\small$\sqrt3$}\,\pi^0 & \mbox{\small$\sqrt6$}^{\vphantom{|}}\,\pi^+ & \mbox{\small$\sqrt6$}\,K^+ \\ \mbox{\small$\sqrt6$}\,\pi^- & \!\!\eta_8^{}\!-\!\mbox{\small$\sqrt3$}\,\pi^0 & \mbox{\small$\sqrt6$}\,K^0 \\ \mbox{\small$\sqrt6$}\,K^- & \mbox{\small$\sqrt6$}~\overline{\!K}{}^0 & -2\eta_8^{} \end{array}\!\right) ,
\nonumber \\
{\cal V}^\mu & \,=\, \frac{1}{2}\big(\xi\,\partial^\mu\xi^\dagger + \xi^\dagger\,\partial^\mu\xi\big) + \frac{i\tt e}{2} \big(\xi^\dagger{\cal Q}_q\;\!\xi + \xi{\cal Q}_q\;\!\xi^\dagger\big) A^\mu , ~~~ ~ \overline{\!B} \,=\, B^\dagger \gamma_0^{} \,, ~~~ ~ \hat\Sigma \,=\, \xi^2 \,=\, e^{i\varphi/2f\!_\pi^{}} , ~~~
\nonumber \\
{\cal A}^\mu & \,=\, \frac{i}{2}\big(\xi\,\partial^\mu\xi^\dagger - \xi^\dagger\,\partial^\mu\xi\big)
+ \frac{\tt e}{2} \big(\xi^\dagger{\cal Q}_q\;\!\xi - \xi{\cal Q}_q\;\!\xi^\dagger\big) A^\mu , ~~~ ~~ {\mathbb D}^\mu\hat\Sigma \,=\, \partial^\mu\hat\Sigma \!+\! \big[{\cal Q}_q,\hat\Sigma\big] i{\tt e}A^\mu \,,
\end{align}
with \,$f\!_\pi^{}=(92.07\pm0.85)$\,MeV\, being the pion decay constant \cite{ParticleDataGroup:2026mpi} and \,${\cal Q}_q={\rm diag}(2,-1,-1)/3$  the light quarks' charge matrix.\footnote{Under the chiral-symmetry group \,${\rm SU}(3)_L\times{\rm SU}(3)_R$\, the hadron fields transform according to \,\,$\overline{\!B}\to\hat U\,\overline{\!B}\hat U^\dagger$,\, $B\to\hat UB\hat U^\dagger$,\, $\hat\Sigma\to\hat L\hat\Sigma\hat R^\dagger$,\, $\xi\to\hat L\xi\hat U^\dagger=\hat U\xi\hat R^\dagger$, \,${\cal V}^\mu\to\hat U{\cal V}^\mu\hat U^\dagger+i\partial^\mu\hat U\,\hat U^\dagger$,\, and \,${\cal A}^\mu\to \hat U{\cal A}^\mu\hat U^\dagger$,\, where \,$\hat U\in{\rm SU}(3)$\, is implicitly defined by the $\xi$ equation, \,$\hat L\in{\rm SU}(3)_L$,  and \,$\hat R\in{\rm SU}(3)_R$.}

After working on eq.\,(\ref{unitarity}) in view of eq.\,(\ref{B2B'g}), we get
\begin{align}
{\rm Im}\,a_\Lambda^{} & = \frac{\cal D{+}F}{4\sqrt2} \Big[ {\cal C}_{pp}^{} \!-\! {\cal C}_{pp}^\prime + \Big( {\cal C}_p^\prime \!-\! {\cal C}_p^{} \Big) \hat u_-^{} + \Big( {\cal C}_{pq}^{} \!-\! {\cal C}_{pq}^\prime \!+\! {\cal C}_q^\prime \Big) \hat v\;\!\!_-^{} \Big] \frac{m_N^{} m_\Lambda^{} m_{\pi^+}^2 {\mathbb B}_{\Lambda p\pi^-}^{}}{\pi^2f} \,,
\nonumber \\
{\rm Im}\,b_\Lambda^{} & = \frac{\cal D{+}F}{4\sqrt2} \Big[ {\cal C}_{pp}^{} \!-\! {\cal C}_{pp}^\prime + \Big( {\cal C}_p^\prime \!-\! {\cal C}_p^{} \Big) \hat u_+^{} + \Big( {\cal C}_{pq}^{} \!-\! {\cal C}_{pq}^\prime \!+\! {\cal C}_q^\prime \Big) \hat v\;\!\!_+^{} \Big] \frac{m_N^{} m_\Lambda^{} m_{\pi^+}^2 {\mathbb A}_{\Lambda p\pi^-}^{}}{\pi^2f} \,,
\nonumber \\
{\rm Im}\,c_\Lambda^{} & = \frac{\cal D{+}F}{4\sqrt2} \Bigg[ \! \begin{array}[t]{l}
\displaystyle {\cal C}\;\!\!_{pq}^{} - {\cal C}_{\;\!\!pq}^\prime - {\cal C}\;\!\!_q^{}
- {\cal C}_{\;\!\!q}^\prime + 2~\!\!{\cal C}\;\!\!_{qq}^{} - 2~\!\!{\cal C}_{\;\!\!qq}^\prime - \frac{ {\cal C}\;\!\!_p^{} \!-\! {\cal C}\;\!\!_{pp}^{} \!+\! {\cal C}_{\;\!\!pp}^\prime \!-\! 2~\!\!{\cal C}\;\!\!_{pq}^{} \!+\! 2~\!\!{\cal C}_{\;\!\!pq}^\prime }{\hat v\;\!\!_-^{}}
\\ \displaystyle \!-\; \Big( {\cal C}\;\!\!_p^{} - {\cal C}_{\;\!\!p}^\prime + 2~\!\!{\cal C}\;\!\!_q^{} - 2~\!\!{\cal C}_{\;\!\!q}^\prime - {\cal J}_0^{} \Big)
\frac{\hat u_-^{}}{\hat v\;\!\!_-^{}} \Bigg] \frac{m_N^{} m_{\pi^+}^2 {\mathbb B}_{\Lambda p\pi^-}^{}}{\pi^2f} \,, \end{array}
\nonumber \\
{\rm Im}\,d_\Lambda^{} & = \frac{\cal D{+}F}{4\sqrt2} \Bigg[ \! \begin{array}[t]{l}
\displaystyle {\cal C}_{\;\!\!pq}^\prime - {\cal C}\;\!\!_{pq}^{} + {\cal C}\;\!\!_q^{} + {\cal C}_{\;\!\!q}^\prime - 2~\!\!{\cal C}\;\!\!_{qq}^{}
+ 2~\!\!{\cal C}_{\;\!\!qq}^\prime + \frac{ {\cal C}\;\!\!_p^{} \!-\!{\cal C}\;\!\!_{pp}^{} \!+\! {\cal C}_{\;\!\!pp}^\prime \!-\! 2~\!\!{\cal C}\;\!\!_{pq}^{}
\!+\! 2~\!\!{\cal C}_{\;\!\!pq}^\prime }{\hat v\;\!\!_+}
\\ \displaystyle \!\!+\; \Big( {\cal C}\;\!\!_p^{} - {\cal C}_{\;\!\!p}^\prime
+ 2~\!\!{\cal C}\;\!\!_q^{} - 2~\!\!{\cal C}_{\;\!\!q}^\prime - {\cal J}_0^{} \Big)
\frac{\hat u_+^{}}{\hat v\;\!\!_+} \Bigg] \frac{m_N^{} m_{\pi^+}^2 {\mathbb A}_{\Lambda p\pi^-}^{}}{\pi^2f} \,, \end{array}
\end{align}
which contain various quantities, namely
\begin{align}
{\cal C}_p^{} & \,=\, \frac{ {\cal J}_0^{~} \hat{\tt y}
- {\cal C}_q^{~} \hat{\tt x} }{2} \,, ~~~ ~~~~
{\cal C}_{pp}^{} \,=\, \frac{12_{\,\!} \mathscr C_g^{~} \hat{\tt r}\;\!\!_q^{}}
{m_\Lambda^2\,\hat{\tt z}} - {\cal I}_0^{}\;\! \frac{ 2~\!\!\hat{\tt r}\;\!\!_q^{}
+ \hat{\tt x}~\!\!\hat{\tt y} }{2\;\!m_\Lambda^2\;\!\hat{\tt z}} + {\cal J}_0^{}\;\!
\hat{\tt r}\;\!\!_q^{}\frac{\hat{\tt r}\;\!\!_q^{}-4~\!\!\hat{\tt r}\;\!\!_\pi^{}}{\hat{\tt z}} \,,
\nonumber \\
{\cal C}_q^{} & \,=\, \frac{2\;\!{\cal I}_0^{}}{m_\Lambda^2\;\! \hat{\tt z}}
+ {\cal J}_0^{}\;\! \frac{\hat{\tt x}~\!\!\hat{\tt y}-2~\!\!\hat{\tt r}\;\!\!_q}{\hat{\tt z}} \,,
~~~ ~~~~ {\cal C}_{pq}^{} \,=\, \frac{ {\cal I}_0^{}\;\! \hat{\tt y}
- 6_{\,\!}\mathscr C_g^{~} \hat{\tt x} }{m_\Lambda^2\,\hat{\tt z}}
+ {\cal J}_0^{}\;\! \frac{ 2~\!\!\hat{\tt r}\;\!\!_\pi^{}\hat{\tt x}
- \hat{\tt r}\;\!\!_q\hat{\tt y} }{\hat{\tt z}} \,,
\nonumber \\
{\cal C}_{qq}^{} & \,=\, \frac{12_{\,\!}\mathscr C_{\;\!\!g}^{}}
{m_\Lambda^2\;\!\hat{\tt z}} + {\cal J}_0^{}\;\! \frac{\hat{\tt y}^2-4~\!\!\hat{\tt r}\;\!\!_\pi^{}}
{\hat{\tt z}} \,, ~~~ ~~~~
\mathscr C_g^{} \,=\, \Bigg( {\cal J}_0^{}\;\! \hat{\tt r}\;\!\!_\pi^{} - \frac{ {\cal C}_p^{~}
\hat{\tt y} + {\cal C}_q^{~} \hat{\tt r}\;\!\!_q^{}}{2} \Bigg) \frac{m_\Lambda^2}{2} \,,
\nonumber \\
{\cal C}_p^\prime & \,=\, \frac{ {\cal J}_0^\prime\;\! \hat{\tt y}
\!-\! {\cal C}_q^{\prime\,} \hat{\tt x} }{2} \,, ~~~ ~~~~
{\cal C}_{pp}^\prime \,=\, \frac{12\;\! \mathscr C_{\!g}^{\prime\,} \hat{\tt r}\;\!\!_q^{}}
{m_\Lambda^2\,\hat{\tt z}} + {\cal I}_0^{}\;\! \frac{ 2~\!\!\hat{\tt r}\;\!\!_q^{}
\!-\! 2~\!\!\hat{\tt x} \!-\! \hat{\tt x}~\!\!\hat{\tt y} }{2\;\!m_\Lambda^2\;\!\hat{\tt z}} + {\cal J}_0^\prime\;\! \frac{ (1 \!-\! \hat{\tt r}\;\!\!_n)^2
\!-\! 4~\!\!\hat{\tt r}\;\!\!_\pi^{} \hat{\tt r}\;\!\!_q^{} }{\hat{\tt z}} \,,
\nonumber \\
{\cal C}_q^\prime & \,=\, \frac{2\;\!{\cal I}_0^{}}{m_\Lambda^2\;\! \hat{\tt z}}
+ {\cal J}_0^\prime\;\!  \frac{ 2~\!\!\hat{\tt r}\;\!\!_n^{} - 2
+ \hat{\tt x}~\!\!\hat{\tt y} }{\hat{\tt z}} \,, ~~~ ~~
{\cal C}_{pq}^\prime \,=\, \frac{{\cal I}_0^{~} \hat{\tt y}
- 6\;\! \mathscr C_{\!g}^{\prime\,} \hat{\tt x}}{m_\Lambda^2\,\hat{\tt z}}
+ {\cal J}_0^\prime\;\! \frac{ 2~\!\!\hat{\tt r}\;\!\!_\pi^{} \hat{\tt x}
+ \hat{\tt r}\;\!\!_q^{}\hat{\tt y} - \hat{\tt x}~\!\!\hat{\tt y} }{\hat{\tt z}} \,,
\nonumber \\
{\cal C}_{qq}^\prime & \,=\, \frac{12\;\! \mathscr C_{\!g}^\prime}
{m_\Lambda^2\;\! \hat{\tt z}} + {\cal J}_0^\prime\;\! \frac{ \hat{\tt y}{}^2
- 4~\!\!\hat{\tt r}\;\!\!_\pi^{} }{\hat{\tt z}} \,, ~~~ ~~~~
\mathscr C_{\!g}^\prime \,=\, \Bigg[ {\cal J}_0^\prime\;\! \hat{\tt r}\;\!\!_\pi^{}
- \frac{ {\cal C}_p^\prime\;\! \hat{\tt y} + \big(1-\hat{\tt r}\;\!\!_n^{}\big)
{\cal C}_{\;\!\!q}^\prime }{2} \Bigg] \frac{m_\Lambda^2}{2} \,,
\nonumber \\
\hat u_\pm^{} & \,=\, 1 \pm \frac{m_p}{\raisebox{3pt}{$m_\Lambda^{}$}} \,,
~~~ ~~ \hat v\;\!\!_\pm^{} \,=\, 1 \pm \frac{m_n^{}}{\raisebox{3pt}{$m_\Lambda^{}$}} \,,
~~~ ~~ {\cal J}_0^{} \,=\, \frac{\pi}{4m_\Lambda^{}\;\! |\hat{\mbox{\bf\small\textit q}}|}
{\rm\,\ln} \Bigg( \frac{ q^2 - 2~\!\!\hat{\mbox{\footnotesize\tt E}}_\pi^{}\hat q_0^{}
+ 2 |\hat{\mbox{\bf\scriptsize P}}\;\!\!\!_\pi^{}| |\hat{\mbox{\bf\small\textit q}}| }
{ q^2 - 2~\!\!\hat{\mbox{\footnotesize\tt E}}_\pi^{}\hat q_0^{} - 2
|\hat{\mbox{\bf\scriptsize P}}\;\!\!\!_\pi^{}| |\hat{\mbox{\bf\small\textit q}}| } \Bigg) \,,
\end{align}
with
\begin{figure}[b]\bigskip\centering\includegraphics[width=0.47\textwidth]{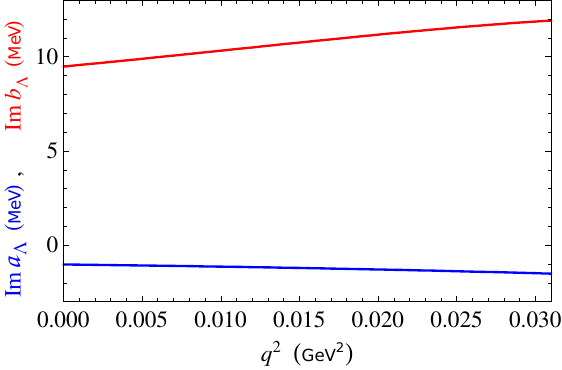} ~~ \includegraphics[width=0.48\textwidth]{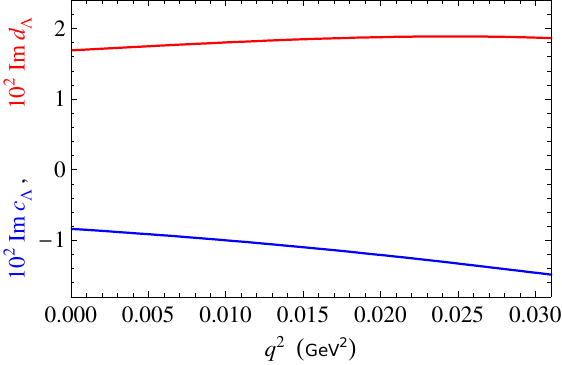}\vspace{-1ex}\caption{The imaginary parts of $a_\Lambda^{}$ and $b_\Lambda^{}$ (left panel) and of $c_\Lambda^{}$ and $d_\Lambda^{}$ (right panel) in the LD amplitude for \,$\Lambda\to n\gamma^*$\, as functions of the photon's squared momentum $q^2$. The $a_\Lambda$ and $c_\Lambda$ $\big(b_\Lambda$ and $d_\Lambda\big)$ curves are in blue (red).} \label{L2ng:im-abcd}
\end{figure}
\begin{align} &
\hat{\tt x} \,=\, 1 - \hat{\tt r}\;\!\!_n^{}
+ \hat{\tt r}\;\!\!_q^{} \,, ~~~ ~~~~
\hat{\tt y} \,=\, 1 - \hat{\tt r}\;\!\!_p^{} + \hat{\tt r}\;\!\!_\pi^{} \,, ~~~ ~~~~
\hat{\tt z} \,=\, \hat{\tt x}{}^2 \!- 4~\!\!\hat{\tt r}\;\!\!_q^{} \,, ~~~ ~~~~
\hat{\tt r}\;\!\!_q^{} \,=\, \frac{q^2}{m_\Lambda^2} \,,
\nonumber \\ &
\hat{\tt r}_{n,\;\!p,\;\!\pi}^{} \,=\, \frac{m_{n,\;\!p,\;\!\pi^+}^2}{m_\Lambda^2} \,, ~~~ ~~~~
{\cal I}_0^{} \,=\, \frac{\pi|\hat{\mbox{\bf\scriptsize P}}\;\!\!\!_\pi^{}|}{m_\Lambda^{}} \,,
~~~ ~~~~ |\hat{\mbox{\bf\scriptsize P}}\;\!\!\!_\pi^{}| \,=\, \frac{m_\Lambda^2}{2} \mbox{$
\sqrt{\hat u_+^2 - \hat{\tt r}\;\!\!_\pi^{}}\,\sqrt{\hat u_-^2 - \hat{\tt r}\;\!\!_\pi^{}}$} \,,
\nonumber \\ &
{\cal J}_0^\prime \,=\, \frac{\pi}{ 4m_\Lambda^{}\;\!
|\hat{\mbox{\bf\small\textit q}}| } {\rm\,\ln} \Bigg( \frac{ m_\Lambda^2 - m_n^2
- 2~\!\!\hat{\mbox{\footnotesize\tt E}}_\pi^{}\hat q_0^{}
+ 2 |\hat{\mbox{\bf\scriptsize P}}\;\!\!\!_\pi^{}||\hat{\mbox{\bf\small\textit q}}| }
{ m_\Lambda^2 - m_n^2 - 2~\!\!\hat{\mbox{\footnotesize\tt E}}_\pi^{}\hat q_0^{} - 2
|\hat{\mbox{\bf\scriptsize P}}\;\!\!\!_\pi^{}| |\hat{\mbox{\bf\small\textit q}}| } \Bigg) \,,
~~~~~ ~~~~~ \hat{\mbox{\footnotesize\tt E}}{}_\pi^{} \,=\, \mbox{$\sqrt{ m_{\pi^+}^2 + \hat{\mbox{\bf\scriptsize P}}{}_{\!\pi}^2}$} \,, ~~~ ~~
\nonumber \\ &
|\hat{\mbox{\bf\small\textit q}}| \,=\, \frac{m_\Lambda^{}}{2} \mbox{$
\sqrt{\hat v_+^2 - \hat{\tt r}\;\!\!_q}\,\sqrt{\hat v_-^2 - \hat{\tt r}\;\!\!_q}$} \,,
~~~~~ ~~~~~ ~~
\hat q_0^{} \,=\, \sqrt{q^2 + \hat{\mbox{\bf\small\textit q}}{}^2} \,.
\end{align}
Numerically, from the measurements~\cite{ParticleDataGroup:2026mpi} on the rate and $\alpha$ parameter of \,$\Lambda\to p\pi^-$\, we infer \,${\mathbb A}_{\Lambda p\pi^-}=1.3881\pm0.0071$\, and \,${\mathbb B}_{\Lambda p\pi^-}=11.63\pm0.16$.\,
In figure \ref{L2ng:im-abcd} we display how these form factors behave over the whole kinematical range, \,$4m_e^2\le q^2\le(m_\Lambda-m_n)^2\simeq0.031\rm\,GeV^2$.\,

As for the real parts of $a_\Lambda^{}$ and $b_\Lambda^{}$, since figure \ref{L2ng:im-abcd} reveals that Im\,$a_\Lambda^{}$ and Im\,$b_\Lambda^{}$ are mildly dependent on $q^2$, whose range is quite narrow, we can reasonably make the approximations \,Re\,$a_\Lambda^{}(q^2)={\rm Re}\,a_\Lambda^{}(0)$\, and \,Re\,$b_\Lambda^{}(q^2)={\rm Re}\,b_\Lambda^{}(0)$,\, where the right-hand sides can be obtained from the empirical information on \,$\Lambda\to n\gamma$\, after incorporating the calculated Im\,$a_\Lambda^{}(0)$ and Im\,$b_\Lambda^{}(0)$.
Using again \,$\big\langle{\cal B}_{\Lambda n\gamma}^{\tt exp}\big\rangle=(8.35\pm0.65)\!\times\!10^{-4}$\, and \,$\big\langle\alpha_{\Lambda n\gamma}^{\tt exp}\big\rangle=-0.16\pm0.10$,\, from eqs.\,\,(\ref{BL2ng-data}) and (\ref{aL2ng-data}), we then find
\begin{align}
{\rm Re}\,a_\Lambda^{} \,=\, (5.75\pm0.87)\rm\,MeV \,, & & {\rm Re}\,b_\Lambda^{} \,=\, (-0.03\pm1.09)\rm\,MeV \,. & ~~~ ~~~~
\end{align}

\begin{figure}[b]\bigskip\centering\includegraphics[width=0.9\textwidth]{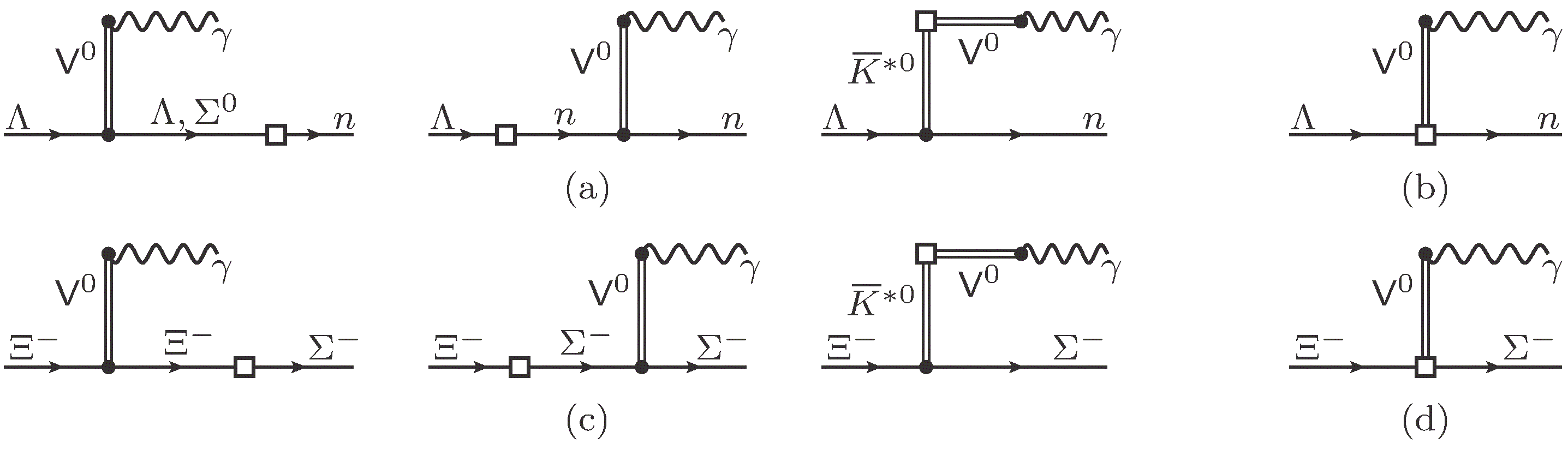}\vspace{-1ex}\caption{Diagrams giving rise to (a)\,Re\,$c_\Lambda^{}$, (b)\,Re\,$d_\Lambda^{}$, (c)\,Re\,$c_\Xi^{}$, and (d)\,Re\,$d_\Xi^{}$, where a single line (double line) represents a baryon (vector meson), a solid dot (hollow square) symbolizes a strong (weak) vertex from ${\cal L}_{\rm s}'$ $({\cal L}_{\rm w}$), and \,$\textsf V^0=\rho^0,\omega,\phi$.}   \label{recred-diagrams}
\end{figure}

To estimate Re\,$c_\Lambda^{}$ and Re\,$d_\Lambda^{}$, we apply the assumption of vector-meson dominance, in the same procedure as that implemented in ref.\,\cite{He:2005yn} for \,$\Sigma^+\to p\gamma^*$.\,
Thus, adopting ideal $\omega$-$\phi$ mixing, we collect the vector-meson fields into
\begin{equation} \label{vfields}
\textsf V \,=\, \frac{1}{\mbox{\small2}}\! \left(\!\begin{array}{ccc} \omega+\rho^0 & \mbox{\small$\sqrt2$}^{\vphantom{|}}\,\rho^+ & \mbox{\small$\sqrt2$}\,K^{*+} \\ \mbox{\small$\sqrt2$}\,\rho^- & \omega-\rho^0 & \mbox{\small$\sqrt2$}\,K^{*0} \\ \mbox{\small$\sqrt2$}\,K^{*-} & \mbox{\small$\sqrt2$}~\overline{\!K}{}^{*0} & \sqrt2\,\phi \end{array}\!\right) . ~~~ ~~
\end{equation}
Their strong and weak interactions with the other hadrons and the photon are described by the effective Lagrangians~\cite{Bijnens:1985kj,Ecker:1989yg,Kubis:2000zd,He:2005yn}
\begin{align} \label{Ls'}
{\cal L}_{\rm s}'\, & =\, {\rm Tr} \big( {\cal G}\,\!\!_D^{} \;\overline{\!B} \gamma^\mu \big\{{\sf V}_\mu,B\big\} + {\cal G}\,\!\!_F^{} \;\overline{\!B} \gamma^\mu \big[{\sf V}_\mu,B\big] \big) + {\cal G}_0^{} {\rm Tr} \big(\;\!\overline{\!B}\gamma^\mu B\big) {\rm Tr} \big({\sf V}_\mu\big)
\nonumber \\ & ~~~ -\, \tfrac{1}{2} {\tt e} f_{\tt V}^{\,~} {\rm Tr} \Big[ \Big(\xi^\dagger{\cal Q}_q^{}\;\!\xi+\xi{\cal Q}_q^{}\;\!\xi^\dagger\Big) \big({\mathbb D}_\mu{\sf V}\!_\tau^{}-{\mathbb D}_\tau^{}{\sf V}_\mu\big) \Big] \big(\partial^\mu \! A^\tau-\partial^\tau\! A^\mu\big) \,, ~~~ ~~~~
\\ \raisebox{3ex}{}
{\cal L}_{\rm w}^{}\, & =\, G_{\rm F}^{}m_{\pi^+}^2 {\rm Tr} \big( h\;\!\!_D^{} \;\overline{\!B} \{\overline\kappa, B\} + h\;\!\!_F^{} \;\overline{\!B} [\overline\kappa, B] + h_{\tt V}^{}\;\! \overline\kappa\;\! \textsf V^\mu\textsf V_\mu + h_{\tt PV}^{}\;\! \overline\kappa\;\! \big\{ \!\big[\,\overline{\!B}, \gamma^\mu\gamma_5^{}B\big], \!\textsf V_\mu\big\} \big) ~~~
\nonumber \\ & ~~~ +\, \rm H.c. \,,
\end{align}
where $f_{\tt V}^{}$, ${\cal G}_{D,F,0}^{}$, and $h_{D,F,\tt V,PV}^{}$ are constants, \,$\overline\kappa=\xi^\dagger h\xi$\, contains a 3$\times$3-matrix $h$ with elements \,$h_{kl}^{}=\delta_{k2}^{}\delta_{3l}^{}$\, selecting out \,$s\to d$\, transitions, and \,${\mathbb D}_\kappa{\sf V}_\tau=\partial_\kappa{\sf V}_\tau+[{\cal V}_\kappa,{\sf V}_\tau]$.\footnote{Under chiral symmetry \,${\sf V}\to\hat U{\sf V}\hat U^\dagger$\, and
\,${\mathbb D}^\mu{\sf V}^\nu\to\hat U{\mathbb D}^\mu{\sf V}^\nu\hat U^\dagger$.}
Accordingly, Re\,$c_\Lambda^{}$ and Re\,$d_\Lambda^{}$ arise from the contributions depicted by the diagrams (a) and (b) in figure~\ref{recred-diagrams}.
From them, with \,$\hat s\equiv q^2$,\, we arrive at
\begin{align}
{\rm Re}\,c_\Lambda^{}\, & = \! \begin{array}[t]{l} \displaystyle \Bigg[ \Bigg( 9\;\! \frac{{\cal G}\!_D^{}\!+\!{\cal G}\!_F^{}}{m_\rho^2\!-\!\hat s} - \frac{{\cal G}\!_D^{}\!+\!3{\cal G}\!_F^{}}{m_\omega^2-\hat s} - 2\;\! \frac{{\cal G}\!_D^{}\!+\!3{\cal G}\!_F^{}}{m_\phi^2-\hat s} \Bigg) \frac{h_D^{}\!+\!3h_F^{}}{m_n^{}\!-\!m_\Lambda^{}} + \frac{18\;\! {\cal G}\!_D^{}}{m_\rho^2\!-\!\hat s} \Bigg(\! \frac{h_D^{}-h_F^{}}{m_n^{}\!-\!m_{\Sigma^0}^{}} \!\Bigg)
\\ \displaystyle \;+\; \frac{3}{2} \Bigg(\frac{3}{m_\rho^2-\hat s} - \frac{1}{m_\omega^2-\hat s} + \frac{2}{m_\phi^2-\hat s}\Bigg) \frac{{\cal G}_D^{}+3{\cal G}_F^{}}{\hat s-m_{K^*}^2}\;\! h_{\tt V}^{} \Bigg] \frac{f_{\tt V}^{}m_{\pi^+}^2}{18\sqrt6} \,, \end{array} ~~~
\nonumber \\ \label{redL}
{\rm Re}\,d_\Lambda^{}\, & =\, \Bigg(\frac{-3}{m_\rho^2-\hat s} + \frac{1}{m_\omega^2-\hat s} - \frac{2}{m_\phi^2-\hat s}\Bigg) \frac{f_{\tt V}^{}\;\!h_{\tt PV}^{}\;\!m_{\pi^+}^2}{2\sqrt6} \,.
\end{align}

\begin{figure}[h]\bigskip\centering\includegraphics[width=0.5\textwidth]{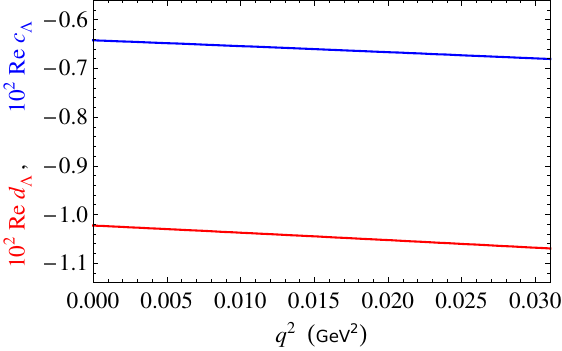}\vspace{-1ex}\caption{The real parts of $c_\Lambda^{}$ and $d_\Lambda^{}$ in the long-distance amplitude for \,$\Lambda\to n\gamma^*$\, versus the photon's squared momentum $q^2$.
The $c_\Lambda$ and $d_\Lambda$ curves are in blue and red, respectively.} \label{L2ng:re-cd}
\end{figure}

Numerically, similarly to ref.\,\,\cite{He:2005yn}, in ${\cal L}_{\rm s}'$ we have \,$f_{\tt V}^{}=0.202$\, from measured \,$\rho^0\to e^+ e^-$\, rate~\cite{ParticleDataGroup:2026mpi} and \,${\cal G}\,\!\!_D^{}=-13.9$,\, ${\cal G}\,\!\!_F^{}=17.9$,\, and \,${\cal G}\,\!\!_0^{}=18.9$\, deduced from refs.\,\,\cite{Kubis:2000zd,Mergell:1995bf}.
Moreover, ${\cal L}_{\rm w}$ includes \,$h_D^{}=-76$\,MeV\, and \,$h_F^{}=179$\,MeV\, from fitting at tree level to the empirical {\tt S}-wave nonleptonic hyperon decays~\cite{ParticleDataGroup:2026mpi,AbdEl-Hady:1999llb}, and \,$h_{\tt V}^{}=-0.34{\rm\,GeV}^2$\, and \,$h_{\tt PV}^{}=2.41$\, extracted from ref.~\cite{Dubach:1996dg}.
The resulting graphs of Re\,$c_\Lambda^{}$ and Re\,$d_\Lambda^{}$ are shown in figure \ref{L2ng:re-cd}.

\section{Form factors in long-distance amplitude for \boldmath$\Xi^-\to\Sigma^-\gamma^*$\label{X2Sg-ff}}

\begin{figure}[t]\centering\includegraphics[width=0.47\textwidth]{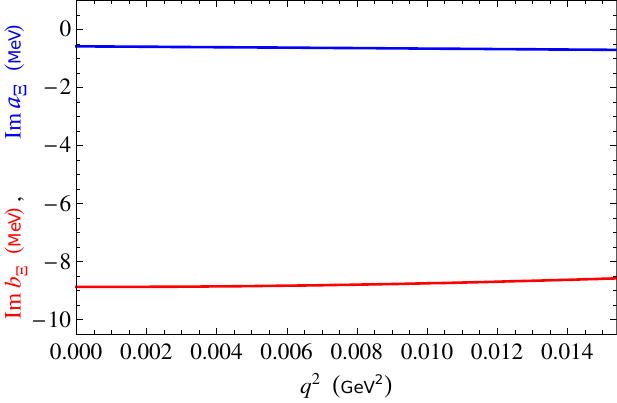} ~~ \includegraphics[width=0.48\textwidth]{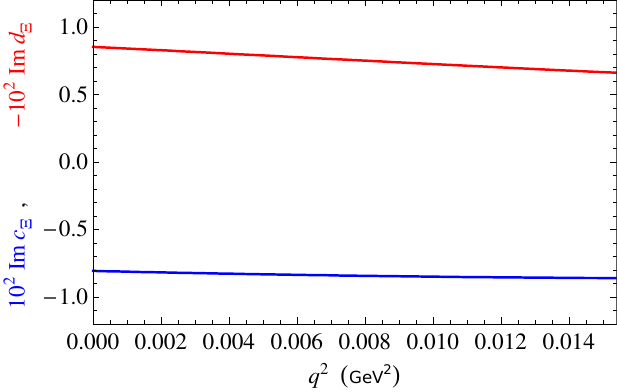}\vspace{-1ex}\caption{The imaginary parts of $a_\Xi^{}$ and $b_\Xi^{}$ (left panel) and of $c_\Xi^{}$ and $-d_\Xi^{}$ (right panel) in the long-distance amplitude for \,$\Xi^-\to\Sigma^-\gamma^*$\, as functions of the photon's squared momentum $q^2$. The $a_\Xi$ and $c_\Xi$ $\big(b_\Xi$ and $d_\Xi\big)$ curves are in blue (red).} \label{Xm2Smg:im-abcd}
\end{figure}

Analogously to the preceding section, here the pertinent unitarity relation is
\begin{equation} \label{imMX2Sg}
{\rm-2\,Im}\,{\cal M}_{\Xi^-\to\Sigma^-\gamma^*}^{} \,=\,
\raisebox{-6pt}{\large$\stackrel{\sum}{\mbox{\scriptsize$\Lambda$\tt\,spin}}$}\; \mbox{\Large$\int$}
\frac{ d{\tt^3P}\!_\pi^{}\;\!d{\tt^3P}\!_\Lambda^{}\;\! \delta^4\! \big({\tt P}\!_\pi^{}
{\tt+P}\!_\Lambda^{}{\tt-P}\!_\Xi^{} \big) }{{\tt16\pi^2\,E}_\pi^{}{\tt\,E}_\Lambda^{}}\,
{\cal M}_{\Lambda\pi\!^-\to\Sigma^-\gamma^*}^{}\;\! {\cal M}_{\Xi^-\to\Lambda\pi\!^-}^{} \,,
\end{equation}
which involves the amplitudes for \,$\Lambda\pi^-\to\Sigma^-\gamma$\, and \,$\Xi^-\to\Lambda\pi^-$\, given by, respectively,
\begin{align} \raisebox{3ex}{}
{-i\cal M}_{\Lambda\pi\!^-\to\Sigma^-\gamma^*}\, & = ~ \raisebox{-7pt}{\includegraphics{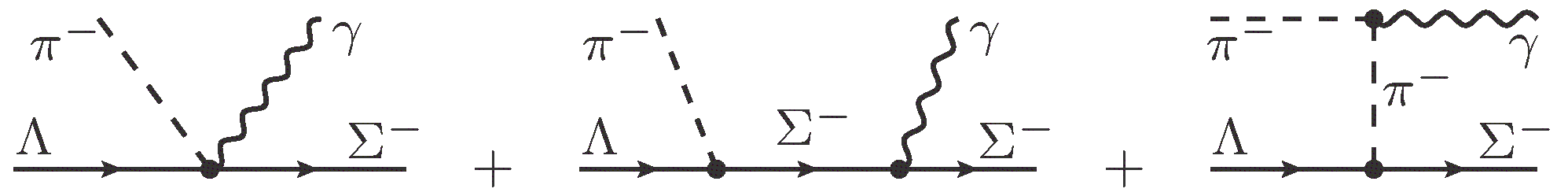}}
\nonumber \\ \raisebox{5ex}{} & =\, \frac{m_\Lambda^{}{+}m_\Sigma^{}}{\sqrt3\,f}\, {\cal D} {\tt\,e}\,\bar u_\Sigma^{}
\Bigg[ \slashed\varepsilon{}^* \frac{m_\Sigma^{}\!+\!\slashed{\tt P}\!_\Xi^{}}
{m_\Sigma^2\!-\!m_\Xi^2} + \frac{\varepsilon^*\!\!\cdot\!\big(2~\!\!{\tt P}\!_\pi^{}\!-\!q\big)}
{2~\!\!{\tt P}\!_\pi^{}\!\cdot q\!-\!q^2} \Bigg] \gamma_5^{} u_\Lambda^{} \,, ~~~ ~~
\end{align}
with \,$m_\Sigma^{}=m_{\Sigma^-}$\, and \,$m_\Xi^{}=m_{\Xi^-}$,\, and \,${\cal M}_{\Xi^-\to\Lambda\pi^-} = i G_{\rm F}^{}m_{\pi^+}^2 \bar u_\Lambda^{} \big( {\mathbb A}_{\Xi^-\Lambda\pi^-} - \gamma_5^{} {\mathbb B}_{\Xi^-\Lambda\pi^-} \big) u_\Xi^{}$.
It follows that
\begin{align}
{\rm Im}\,a_\Xi^{} & \,=\, \frac{m_\Lambda^{}{+}m_\Sigma^{}}{8\sqrt3\,\pi^2f} \bigg( \bar{\tt C}_{pp}^{} - \bar{\tt C}_p^{}\;\! \frac{m_\Xi^{}\!-\!m_\Lambda^{}}{\raisebox{3pt}{$m_\Xi^{}$}} + \bar{\tt C}_{pq}^{}\;\! \frac{m_\Xi^{}\!-\!m_\Sigma^{}}{\raisebox{3pt}{$m_\Xi^{}$}} \bigg) m_\Xi^{~\;} m_{\pi^+}^2 {\mathbb B}_{\Xi^-\Lambda\pi^-}^{}\cal D \,, ~~~ ~~
\nonumber \\
{\rm Im}\,b_\Xi^{} & \,=\, \frac{m_\Lambda^{}{+}m_\Sigma^{}}{8\sqrt3\,\pi^2f} \bigg( \bar{\tt C}_{pp}^{} - \bar{\tt C}_p^{}\;\! \frac{m_\Xi^{}\!+\!m_\Lambda^{}}{\raisebox{3pt}{$m_\Xi^{}$}} + \bar{\tt C}_{pq}^{}\;\! \frac{m_\Xi^{}\!+\!m_\Sigma^{}}{\raisebox{3pt}{$m_\Xi^{}$}} \bigg) m_\Xi^{~\;} m_{\pi^+}^2 {\mathbb A}_{\Xi^-\Lambda\pi^-}^{}\cal D \,,
\nonumber \\
{\rm Im}\,c_\Xi^{} & \,=\, \frac{m_\Lambda^{}{+}m_\Sigma^{}}{8\sqrt3\,\pi^2f} \bigg[ \! \begin{array}[t]{l} \displaystyle \bar{\tt C}_{pq}^{} + {\tt2_{\,\!}\bar C}_{qq}^{} - \bar{\tt C}_q^{} + \Big(\bar{\tt C}_{pp}^{}+{\tt2_{\,\!}\bar C}_{pq}^{}-\bar{\tt C}_p^{}\Big)
\frac{m_\Xi^{}}{\raisebox{3pt}{$m_\Xi^{}\!-\!m_\Sigma^{}$}}
\\ \displaystyle \!+\; \Big(\,\overline{\tt\!J}_0^{}-\bar{\tt C}_p^{}-{\tt2_{\,\!}\bar C}_q^{}\Big)
\frac{m_\Xi^{}\!-\!m_\Lambda^{}}{\raisebox{3pt}{$m_\Xi^{}\!-\!m_\Sigma^{}$}} \bigg] m_{\pi^+\;\!}^2 {\mathbb B}_{\Xi^-\Lambda\pi^-}^{}\cal D \,, \end{array}
\nonumber \\
{\rm Im}\,d_\Xi^{} & \,=\, \frac{m_\Lambda^{}{+}m_\Sigma^{}}{8\sqrt3\,\pi^2f} \bigg[ \! \begin{array}[t]{l} \displaystyle \bar{\tt C}_q^{} - \bar{\tt C}_{pq}^{}
- {\tt2_{\,\!}\bar C}_{qq}^{} - \Big( \bar{\tt C}_{pp}^{} + {\tt2_{\,\!}\bar C}_{pq}^{}
- \bar{\tt C}_p^{} \Big) \frac{m_\Xi^{}}{\raisebox{3pt}{$m_\Xi^{}\!+\!m_\Sigma^{}$}}
\\ \displaystyle \!-\; \Big(\,\overline{\tt\!J}_0^{}-\bar{\tt C}_p^{}-{\tt2_{\,\!}\bar C}_q^{}\Big)
\frac{m_\Xi^{}\!+\!m_\Lambda^{}}{\raisebox{3pt}{$m_\Xi^{}\!+\!m_\Sigma^{}$}} \bigg]
m_{\pi^+\;\!}^2 {\mathbb A}_{\Xi^-\Lambda\pi^-}^{}\cal D \,, \end{array}
\end{align}
where
\begin{align}
\bar{\tt C}_p^{} & \,=\, \frac{ \,\overline{\tt\!J}_0^{\,~} \bar{\tt y}
- \bar{\tt C}_q^{\,~} \bar{\tt x} }{2} \,, ~~~~~ ~~~~~
\bar{\tt C}_{pp}^{} \,=\, \frac{12\;\!\overline{\tt c}\;\!\!_g^{\,~} \bar{\tt r}\;\!\!_q^{}}
{m_\Xi^2\,\bar{\tt z}} - \overline{\tt I}_0^{}\;\! \frac{ 2_{\,\!}\bar{\tt r}\;\!\!_q + \bar{\tt x}\;\!\bar{\tt y} }{2_{\,\!}m_\Xi^2\,\bar{\tt z}} + \,\overline{\tt\!J}_0^{\,~}
\bar{\tt r}\;\!\!_q^{}\;\! \frac{\bar{\tt r}\;\!\!_q^{}-4~\!\!\bar{\tt r}\;\!\!_\pi^{}}{\bar{\tt z}} \,, ~~~ ~~~~
\nonumber \\
\bar{\tt C}_q^{} & \,=\, \frac{2_{\,\!}\overline{\tt I}_0^{}}{m_\Xi^2\;\!\bar{\tt z}}
+ \,\overline{\tt\!J}_0^{}\;\! \frac{\bar{\tt x}\;\!\bar{\tt y}-2~\!\!\bar{\tt r}\;\!\!_q^{}}{\bar{\tt z}} \,,
~~~~~ ~~~~~ \bar{\tt C}_{pq}^{} \,=\, \frac{ \overline{\tt I}_0^{\,~} \bar{\tt y}
- 6\;\!\overline{\tt c}\;\!\!_g^{\,~} \bar{\tt x}}{m_\Xi^2\,\bar{\tt z}} + \,\overline{\tt\!J}_0^{}\;\!
\frac{2~\!\!\bar{\tt r}\;\!\!_\pi^{}\bar{\tt x} - \bar{\tt r}\;\!\!_q^{}\bar{\tt y}}{\bar{\tt z}} \,,
\nonumber \\
\bar{\tt C}_{qq}^{} & \,=\, \frac{12\;\!\overline{\tt c}\;\!\!_g^{}}{ m_\Xi^2\;\!
\bar{\tt z} } + \,\overline{\tt\!J}_0^{}\;\! \frac{\bar{\tt y}{}^2-4~\!\!\bar{\tt r}\;\!\!_\pi^{}}
{\bar{\tt z}} \,, ~~~~~ ~~~~~
\overline{\tt c}\;\!\!_g^{} \,=\, \,\overline{\tt\!J}_0^{}\;\! \frac{m_\Xi^2\;\!\bar{\tt r}\;\!\!_\pi^{}}{2}
- m_\Xi^2\;\! \frac{\bar{\tt C}_p^{\,~}\bar{\tt y}+\bar{\tt C}_q^{~}\bar{\tt r}\;\!\!_q^{}}{4} \,,
\nonumber \\
\overline{\tt\!J}_0^{} & \,=\, \frac{\pi}{4m_\Xi^{}\;\!
|\bar{\mbox{\bf\textit q}}|} {\rm\,\ln}\Bigg( \frac{ q{\tt^2 - 2~\!\!\bar E_\pi^{}}\bar q_0^{}
+ 2 |\bar{\mbox{\bf\scriptsize P}}\,\!\!_\pi^{}| |\bar{\mbox{\bf\textit q}}| }
{ q{\tt^2 - 2~\!\!\bar E}_\pi^{}\bar q_0^{} - 2 |\bar{\mbox{\bf\scriptsize P}}\,\!\!_\pi^{}|
|\bar{\mbox{\bf\textit q}}| } \Bigg) \,,
\end{align}
with
\begin{align}
\bar{\tt x} & \,=\, 1 - \bar{\tt r}_\Sigma^{}+\bar{\tt r}\;\!\!_q^{} \,, ~~~~~ ~~~~~
\bar{\tt y} \,=\, 1 - \bar{\tt r}_\Lambda^{} + \bar{\tt r}\;\!\!_\pi^{} \,, ~~~~~ ~~~~~
\bar{\tt z} \,=\, \bar{\tt x}{}^2 - 4~\!\!\bar{\tt r}\;\!\!_q^{} \,,
\nonumber \\ &
\bar{\tt r}_{\Sigma,\,\Lambda,\,\pi}^{} \,=\,
\frac{m_{\Sigma\!,\,\Lambda,\,\pi^+}^2}{m_\Xi^2} \,, ~~~~~ ~~~~~
\bar{\tt r}_q^{} \,=\, \frac{q^2}{m_\Xi^2} \,, ~~~~~ ~~~~~
\overline{\tt I}_0^{} \,=\, \frac{\pi|\bar{\mbox{\bf\scriptsize P}}\,\!\!_\pi^{}|}{m_\Xi^{}} \,,
\nonumber \\
|\bar{\mbox{\bf\scriptsize P}}\,\!\!_\pi^{}| & \,=\, \frac{\sqrt{(m_\Xi \!+\! m_\Lambda)^2 - m_{\pi^+}^2}\; \sqrt{(m_\Xi \!-\! m_\Lambda)^2 - m_{\pi^+}^2}}{2m_\Xi^{}} \,, ~~~ ~~~~
{\tt\bar E}_\pi^{} \,=\, \mbox{$\sqrt{m_{\pi^+}^2 + \bar{\mbox{\bf\scriptsize P}}_{\,\!\!\pi}^2}$} \,, ~~~ ~~
\nonumber \\
|\bar{\mbox{\bf\textit q}}| & \,=\, \frac{\sqrt{(m_\Xi \!+\! m_\Sigma)^2 - q^2}\; \sqrt{(m_\Xi \!-\! m_\Sigma)^2 - q^2}}{2m_\Xi^{}} \,, ~~~ ~~~~
\bar q_0^{} \,=\, \sqrt{q^2 + \bar{\mbox{\bf\textit q}}^2} \,.
\end{align}
The empirical information~\cite{ParticleDataGroup:2026mpi} on \,$\Xi^-\to\Lambda\pi^-$\, implies that \,$\mathbb A_{\Xi^-\Lambda\pi^-}=-1.9935\pm0.0091$\, and \,$\mathbb B_{\Xi^-\Lambda\pi^-}=6.343\pm0.054$.\,
In figure \ref{Xm2Smg:im-abcd} we graph the $q^2$ dependence of these form factors over the whole kinematical range, \,$4m_e^2\le q^2\le(m_{\Xi^-}-m_{\Sigma^-}){}^2$.\,

For the real parts of $a_\Xi^{}$ and $b_\Xi^{}$, analogously to the preceding \,$\Lambda\to n\gamma^*$\, case, we make the approximations \,Re\,$a_\Xi^{}(q^2)={\rm Re}\,a_\Xi^{}(0)\equiv{\rm Re}\,a_\Xi^{}$\, and \,Re\,$b_\Xi^{}(q^2)={\rm Re}\,b_\Xi^{}(0)\equiv{\rm Re}\,b_\Xi^{}$,\, where the right-hand sides are fixed from the data in eq.\,(\ref{X2Sg-data}) after incorporating the computed Im\,$a_\Xi^{}(0)$ and Im\,$b_\Xi^{}(0)$.
Thus, letting Re\,$a_\Xi^{}$ and Re\,$b_\Xi^{}$ vary freely under the condition that they respect the 1$\sigma$ range of $\mathscr B_{\Xi^-\Sigma^-\gamma}^{\tt exp}$, we arrive at \,$-0.06 < \alpha_{\Xi^-\Sigma^-\gamma} < 0.28$,\, which is compatible with $\alpha_{\Xi^-\Sigma^-\gamma}^{\tt exp}$ and may be regarded as a prediction to be tested with more precise measurements in the future.
Moreover, we can pick
\begin{align}
{\rm Re}\,a_\Xi^{} \,=\, (1.1 \pm 1.1)\rm\,MeV \,, & & {\rm Re}\,b_\Xi^{} \,=\, (1.1 \pm 1.1)\rm\,MeV \,, & ~~~ ~~~~
\end{align}
in light of $\alpha_{\Xi^-\Sigma^-\gamma}^{\tt exp}$ having a huge uncertainty,
which satisfy the different requirements.

\begin{figure}[b]\bigskip\centering\includegraphics[width=0.5\textwidth]{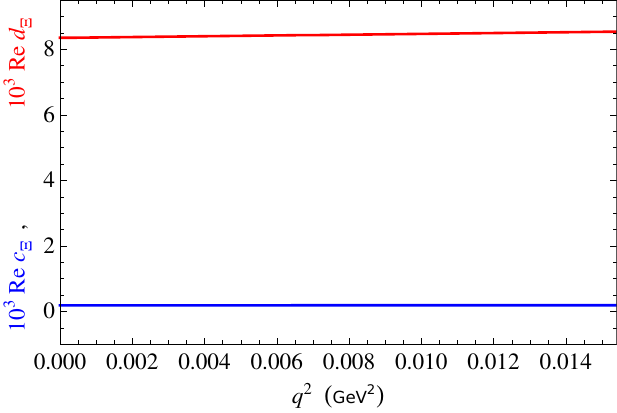}\vspace{-1ex}\caption{The real parts of $c_\Xi$ and $d_\Xi$ in the long-distance amplitude for \,$\Xi^-\to\Sigma^-\gamma^*$\, versus the photon's squared momentum $q^2$.
The $c_\Xi$ and $d_\Xi$ curves are in blue and red, respectively.} \label{Xm2Smg:re-cd}
\end{figure}

To estimate Re\,$c_\Xi^{}$ and Re\,$d_\Xi^{}$, we again use the vector-meson dominance model.
Thus, they proceed from the contributions depicted by diagrams (c) and (d) in figure~\ref{recred-diagrams}, leading to
\begin{align}
{\rm Re}\,c_\Xi^{}\, & = \! \begin{array}[t]{l} \displaystyle \Bigg[ \Bigg( \frac{3}{m_\rho^2\!-\!\hat s} - \frac{1}{m_\omega^2\!-\!\hat s} - \frac{2}{m_\phi^2\!-\!\hat s} \Bigg) \frac{2\big({\cal G}_D^{}+{\cal G}_F^{}\big)}{m_{\Xi^-}^{}-m_{\Sigma^-}^{}} \big(h_D^{}+h_F^{}\big)
\\ \displaystyle \;+\; \Bigg(\frac{3}{m_\rho^2\!-\!\hat s} - \frac{1}{m_\omega^2\!-\!\hat s} + \frac{2}{m_\phi^2\!-\!\hat s}\Bigg) \frac{{\cal G}_D^{}+{\cal G}_F^{}}{m_{K^*}^2-\hat s}\;\! h_{\tt V}^{} \Bigg] \frac{f_{\tt V}^{} m_{\pi^+}^2}{12} \,, \end{array}
\nonumber \\
{\rm Re}\,d_\Xi^{}\, & =\, \frac{-\sqrt2\,\,{\rm Re}\,d_\Lambda^{}}{\sqrt3} \,.
\end{align}
Their relationships to \,$\hat s=q^2$\, are exhibited in figure \ref{Xm2Smg:re-cd}.

\section{Imaginary parts of form factors in LD amplitude for \boldmath$\Sigma^+\to p\gamma^*$\label{S2pg-ff}}

For this transition the unitarity relation  is
\begin{align} \begin{array}[b]{rl}
{\rm-2\,Im\,\cal M}_{\Sigma^+\to p\gamma^*}^{} \! & \displaystyle =\,
\raisebox{-6pt}{\large$\stackrel{\sum}{\mbox{\scriptsize$n$\tt\,spin}}$}\; \mbox{\Large$\int$}
\frac{ d{\tt^3P}\!_\pi^{}\;\! d{\tt^3P}\!_n^{}\;\! \delta^4\!\big( {\tt P}\!_\pi^{}
{\tt+P}\!_n^{}{\tt-P}\!_\Sigma^{} \big) }{{\tt16\pi^2\,E}_\pi^{}{\tt\,E}_n^{}}\,
{\cal M}_{n\pi^+\to p\gamma^*}^{}\, {\cal M}_{\Sigma^+\to n\pi^+}^{}
\vspace{2pt} \\ & \displaystyle ~~ + ~
\raisebox{-6pt}{\large$\stackrel{\sum}{\mbox{\scriptsize$p'$\tt\,spin}}$}\; \mbox{\Large$\int$}
\frac{ d{\tt^3P}\!_\pi^{}\;\! d{\tt^3P}\!_{p'}\;\! \delta^4\!\big( {\tt P}\!_\pi^{}
{\tt+P}\!_{p'}{\tt-P}\!_\Sigma^{} \big) }{{\tt16\pi^2\,E}_\pi^{}{\tt\,E}_{p'}}\,
{\cal M}_{p'\pi^0\to p\gamma^*}^{}\, {\cal M}_{\Sigma^+\to p'\pi^0}^{} \,, \end{array}
\end{align}
where the amplitudes inside the integrals are given by
\begin{align} \label{MSigmaNpi+} \raisebox{3ex}{}
{-i\cal M}_{n\pi^+\to p\gamma^*}\, & \begin{array}[t]{l} = ~ \raisebox{-7pt}{\includegraphics{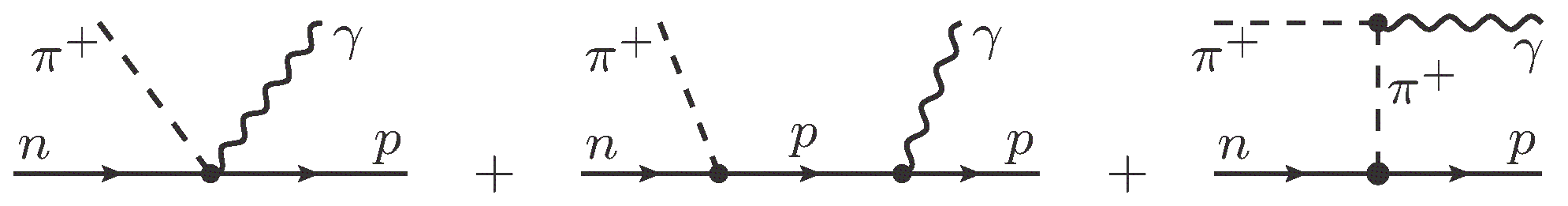}}
\vspace{7pt} \\ \displaystyle =\, \frac{m_p^{}{+}m_n^{}}{\sqrt2\,f} ({\cal D{+}F}){\tt e}\, \bar u_p^{} \Bigg[ \slashed\varepsilon{}^*\, \frac{m_p-m_n^{}+\slashed{\tt P}\!_\pi^{}}{\tilde m_\Sigma^2-m_p^2}
+ \frac{\varepsilon^*\!\cdot\!\big({\tt2P}\!_\pi^{}\!-\!q\big)}
{q{\tt^2-2~\!\!P}\!_\pi^{}\!\!\cdot\!q} \Bigg] \gamma_5^{}u_n^{} \,, \end{array} ~~~ ~~
\nonumber \\ \raisebox{4ex}{}
{-i\cal M}_{\Sigma^+\to n\pi^+}\, & =\, G_{\rm F}^{}m_{\pi^+}^2 \bar u_n^{} \big( \mathbb A_{\Sigma^+n\pi^+} - \gamma_5^{} \mathbb B_{\Sigma^+n\pi^+} \big) u_\Sigma^{} \,,
\end{align}
\begin{align} \label{MSigmaNpi0} \raisebox{3em}{}
{-i\cal M}_{p'\pi^0\to p\gamma^*}^{}\, & \begin{array}[t]{l} = ~ \raisebox{-7pt}{\includegraphics{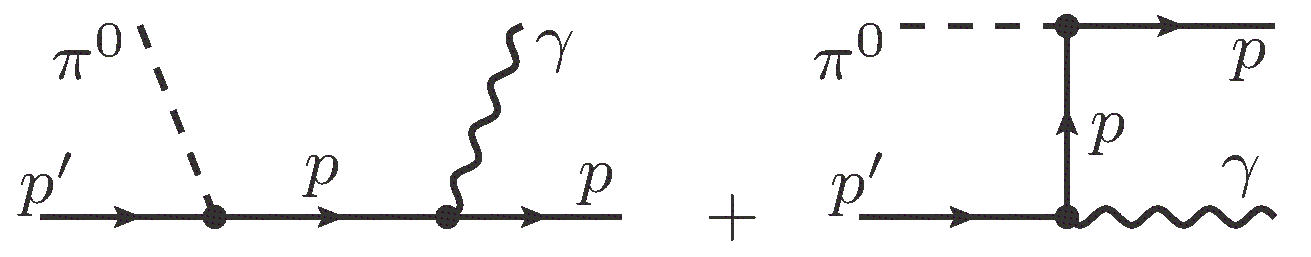}}
\vspace{7pt} \\ \displaystyle =\, \frac{({\cal D{+}F}){\tt e}\,m_p^{}}{f}\, \bar u_p^{} \Bigg( \frac{ \slashed\varepsilon{}^*\, \slashed{\tt P}\!_\pi^{}}{\tilde m_\Sigma^2-m_p^2} + \frac{
\slashed{\tt P}\!_\pi^{}\, \slashed\varepsilon{}^* }{ {\tt2~\!\!P}\!_\pi^{}{\tt\cdot P}\!_p - m_{\pi^0}^2 } \Bigg) \gamma_5^{} u_{p'}^{} \,,
\end{array} ~~~ ~~~~
\nonumber \\ \raisebox{4ex}{}
{-i\cal M}_{\Sigma^+\to p'\pi^0}\, & =\, G_{\rm F}^{}m_{\pi^+}^2 \bar u_{p'}^{} \big(
\mathbb A_{\Sigma^+p\pi^0} - \gamma_5^{} \mathbb B_{\Sigma^+p\pi^0} \big) u_\Sigma^{} \,,
\end{align}
with \,$\widetilde m_\Sigma^{}=m_{\Sigma^+}$.\,
It follows that
\begin{align}
{\rm Im}\,a_\Sigma^{} & = \frac{\cal D{+}F}{4\pi^2f} \Bigg[ \!\! \begin{array}[t]{l} \displaystyle
\frac{m_N^{}}{\sqrt2} \Big( {\tt C}_p^{} {\tt~\!\!u}_-^{} - {\tt C}_{pq}^{} {\tt v\;\!\!_-^{}} - {\tt C}_{pp}^{} \Big) \mathbb B_{\Sigma^+n\pi^+}^{}
\\ \displaystyle \!+\, \frac{m_p}{2} \Big( {\tt C}_{\Sigma\Sigma}' + {\tt C}_{pp}'
{\tt~\!\!r}_{\;\!\!p}^{\tt1/2} - {\tt C}_\Sigma' {\tt v\;\!\!_-^{}} + {\tt C}_{\Sigma p}' {\tt v\;\!\!_+^{}} \Big) \mathbb B_{\Sigma^+p\pi^0}^{} \Bigg] \tilde m_\Sigma^{} m_{\pi^+}^2 \,, \end{array}
\nonumber \\
{\rm Im}\,b_\Sigma^{} & = \frac{\cal D{+}F}{4\pi^2f} \Bigg[ \!\! \begin{array}[t]{l} \displaystyle
\frac{m_N^{}}{\sqrt2} \Big( {\tt C}\;\!\!_p^{} {\tt~\!\!u}_+^{} - {\tt C}\;\!\!_{pq}^{}
{\tt v_+^{}} - {\tt C}_{pp}^{} \Big) \mathbb A_{\Sigma^+n\pi^+}^{}
\nonumber \\ \displaystyle \!+\, \frac{m_p}{2} \Big( {\tt C}_{\Sigma\Sigma}' - {\tt C}_{pp\;\!}' {\tt r}_{\;\!\!p}^{\tt1/2} - {\tt C}_\Sigma' {\tt v\;\!\!_+^{}} + {\tt C}_{\Sigma p}' {\tt v\;\!\!_-^{}} \Big) \mathbb A_{\Sigma^+p\pi^0}^{} \Bigg] \tilde m_\Sigma^{} m_{\pi^+}^2 \,, \end{array}
\\
{\rm Im}\,c_\Sigma^{} & = \frac{\cal D{+}F}{4\pi^2f} \Bigg[ \!\! \begin{array}[t]{l} \displaystyle
\frac{m_N^{}}{\sqrt2} \bigg( {\tt C}_q^{} \!-\! {\tt C}_{pq}^{} {-} {\tt2C}_{qq}^{}
+ \frac{{\tt C}_p\!-\!{\tt C}_{pp}{-}{\tt2C}_{pq}}{\tt v\;\!\!_-^{}} + \frac{ {\tt C}_p
{+} {\tt2C}_q {-} \tt J_0}{\tt v\;\!\!_-^{}} {\tt\;\!u_-^{}} \bigg)\mathbb B_{\Sigma^+n\pi^+}^{}
\\ \displaystyle \!-\, \frac{m_p}{2} \Bigg( {\tt C}_\Sigma' + {\tt C}_{\Sigma p}'
- \frac{{\tt C}_{\Sigma\Sigma}' - {\tt C}_{pp}' {\tt\;\!r}_{\;\!\!p}^{\tt1/2}}{\tt v\;\!\!_-^{}} \Bigg) \mathbb B_{\Sigma^+p\pi^0}^{} \Bigg] m_{\pi^+}^2 \,, \end{array}
\nonumber \\ \label{imdS}
{\rm Im}\,d_\Sigma^{} & = \frac{\cal D{+}F}{4\pi^2f} \!\Bigg[ \!\! \begin{array}[t]{l} \displaystyle
\frac{m_N^{}}{\sqrt2} \bigg( {\tt C}_{pq}^{} \!-\! {\tt C}_q^{} {+} {\tt2C}_{qq}^{}
- \frac{{\tt C}_p\!-\!{\tt C}_{pp} \!-\! {\tt2C}_{pq}}{\tt v\;\!\!_+} - \frac{ {\tt C}_p
{+} {\tt2C}_q {-} \tt J_0}{\tt v\;\!\!_+} {\tt\;\!u_+^{}} \bigg) \mathbb A_{\Sigma^+n\pi^+}^{}
\\ \displaystyle \!+\, \frac{m_p}{2} \Bigg( {\tt C}_\Sigma' + {\tt C}_{\Sigma p}' - \frac{
{\tt C}{}_{\Sigma\Sigma}' + {\tt C}_{pp}' {\tt\;\!r}_{\;\!\!p}^{\tt1/2} }{\tt v\;\!\!_+^{}} \Bigg) \mathbb A_{\Sigma^+p\pi^0}^{} \Bigg] m_{\pi^+}^2 \,, \end{array} ~~~
\end{align}
where
\begin{equation} \!\!\! \begin{array}[b]{rl}
{\tt C}_p^{} \! & \displaystyle =\, \frac{{\tt J_0^{}Y - C}{}_q^{~\,}\tt X}{2} \,, ~~~~~ ~~~~~
{\tt C}_{pp}^{} \,=\, \frac{12\;\!\textsc c\;\!\!_g^{~\,} {\tt r}\;\!\!_q^{}}
{\widetilde m_\Sigma^2\;\!\tt Z} - {\tt I_0^{}}\;\! \frac{{\tt2~\!\!r}\;\!\!_q\tt + X~\!\!Y}
{2_{\,\!}\widetilde m_\Sigma^2\;\!\tt Z} + {\tt J_0^{~\,}r}\;\!\!_q^{}\;\! \frac{ {\tt r}\;\!\!_q
{\tt-4~\!\!r}\;\!\!_\pi^{} }{\tt Z} \,,
\vspace{3pt} \\
{\tt C}_q^{} \! & \displaystyle =\, \frac{\tt2_{\!\,}I_0^{}}{\widetilde m_\Sigma^2\;\! \tt Z}
+ {\tt J_0^{}}\;\! \frac{{\tt X~\!\!Y - 2~\!\!r}\;\!\!_q}{\tt Z} \,, ~~~~~ ~~~~~
{\tt C}_{pq}^{} \,=\, \frac{{\tt I_0^{}Y}-6\;\!\textsc c\;\!\!_g^{~\,}\tt X}{\widetilde m_\Sigma^2\;\!\tt Z}
+ {\tt J_0^{}}\;\! \frac{{\tt2~\!\!r}\;\!\!_\pi^{}{\tt X - r}\;\!\!_q^{} \tt Y}{\tt Z} \,,
\\
{\tt C}_{qq}^{} \! & \displaystyle =\, \frac{12\;\!\textsc c\;\!\!_g^{}}{\widetilde m_\Sigma^2\;\!\tt Z}
+ {\tt J_0^{}}\;\! \frac{{\tt Y^2\!-\!4~\!\!r}\;\!\!_\pi^{}}{\tt Z} \,, ~~~~~ ~~~~~
\textsc c\;\!\!_g^{} \,=\, \Bigg(\! {\tt J_0^{~\,}r}\;\!\!_\pi^{}
- \frac{{\tt C}_p^{}{\tt Y\!+\!C}_q^{~\,}{\tt r}\;\!\!_q^{}}{2} \Bigg) \frac{\widetilde m_\Sigma^2}{2} \,,
\\
{\tt u}_\pm^{} \! & \displaystyle =\, 1 \pm \frac{m_n^{}}{\widetilde m_\Sigma^{}} \,, ~~~ ~~
{\tt v}\;\!\!_\pm^{} \,=\, 1 \pm \frac{m_p}{\widetilde m_\Sigma^{}} \,, ~~~ ~~
{\tt~J_0^{}} \,=\, \frac{\pi}{4\widetilde m_\Sigma^{}\;\! |\mbox{\bf\textit q}|}{\rm\,\ln}
\Bigg( \frac{ q{\tt^2 - 2~\!\!E}_\pi^{} q_0^{} + 2 |\mbox{\bf\footnotesize P}\;\!\!\!_\pi|
|\mbox{\bf\textit q}| }{ q{\tt^2 - 2~\!\!E}_\pi^{} q_0^{}
- 2 |\mbox{\bf\footnotesize P}\;\!\!\!_\pi||\mbox{\bf\textit q}|} \Bigg) ,
\\
\tt X \! & \displaystyle =\, {\tt1 - r}\;\!\!_p^{} + {\tt r}\;\!\!_q^{} \,, ~~~ ~~~~
{\tt Y \,=\, 1 - r}\;\!\!_n^{} + {\tt r}\;\!\!_\pi^{} \,, ~~~ ~~~~
{\tt Z \,=\, X^2 - 4~\!\!r}\;\!\!_q^{} \,, ~~~ ~~~~
{\tt r}\;\!\!_q^{} \,=\, \frac{q^2}{\widetilde m_\Sigma^2} \,,
\\ & \hspace{-3ex} {\displaystyle {\tt r}\;\!\!_{p,n,\pi}^{} \,=\, \frac{m_{p,n,\pi^+}^2}{\widetilde m_\Sigma^2} \,, ~~~~~ ~~~~~ {\tt I_0^{}} \,=\, \frac{\pi|\mbox{\bf\footnotesize P}\;\!\!\!_\pi^{}|}{\widetilde m_\Sigma^{}} \,, ~~~~~ ~~~~~
|\mbox{\bf\footnotesize P}\;\!\!\!_\pi^{}| \,=\, \frac{\widetilde m_\Sigma^{}}{2} }
\sqrt{{\tt u_+^2-r}\;\!\!_\pi^{}}\,\sqrt{{\tt u_-^2-r}\;\!\!_\pi^{}} \,,
\\
{\tt~E}_\pi^{} \! & =\, \sqrt{m_{\pi^+}^2 + \mbox{\bf\footnotesize P}_{\!\pi}^2} \,, ~~~ ~~~~
q_0^{} \,=\, \sqrt{q^2 + \mbox{\bf\textit q}^2} \,, ~~~ ~~~~
|\mbox{\bf\textit q}| \,=\, {\displaystyle\frac{\widetilde m_\Sigma^{}}{2}}
\sqrt{{\tt v_+^2-r}\;\!\!_q}\, \sqrt{{\tt v_-^2-r}\;\!\!_q} \,, \end{array}
\end{equation}
\begin{align} \!\!\! \begin{array}[b]{rl}
{\tt C}_\Sigma' \!\! & \displaystyle = \frac{{\tt J_0'Y'\!\!-\!C}{}_p'\tt X'}{2} \,, ~~~ ~~~
{\tt C}_{\;\!\!\Sigma\Sigma}' = \frac{12_{\,\!}\textsc c_{\;\!\!g}'{\tt r}\;\!\!_p^{}}
{\widetilde m_\Sigma^2\tt\;\!Z} \!-\! {\tt I_0'}\frac{{\tt2~\!\!r}{}_{\!\pi}'\tt{+}X'~\!\!Y'}
{2_{\,\!}\widetilde m_\Sigma^2\tt\;\!Z} \!+\! {\tt J}_0^{\prime~\!}{\tt r}_{\!\pi}'
\frac{{\tt r}{}_{\!\pi}'{\tt-4~\!\!r}_p}{\tt Z} \,, ~~~ ~~~
{\tt X'=1{+}r}\;\!\!_p^{}{\tt-r}\;\!\!_q^{} \,,
\vspace{2pt} \\
{\tt C}_p' \!\! & \displaystyle = \frac{\tt2\;\!I_0'}{\widetilde m_\Sigma^2\tt\;\!Z}
\!+\! {\tt J_0'} \frac{{\tt X'~\!\!Y'\!\!-\!2~\!\!r}{}_{\!\pi}'}{\tt Z} \,, ~~~ ~~~
{\tt C}_{\;\!\!\Sigma p}' = \frac{{\tt I_0'Y' \!\!-\! 6_{\,\!}}\textsc c_{\;\!\!g\,\!}'\tt X'}
{\widetilde m_\Sigma^2\tt\;\!Z} \!+\! {\tt J}_0^{\prime~\!}{\tt r}{}_{\!\pi}'
\frac{\tt2~\!\!X'\!-\!Y'}{\tt Z} \,, ~~~ ~~~ {\tt Y' = 1 \!-\! r}\;\!\!_p^{}\!+\!{\tt r}_{\!\pi}' \,,
\vspace{1pt} \\ & \displaystyle \! \!\!\!
{\tt C}_{\;\!\!pp}' = \frac{12_{\,\!}\textsc c_{\;\!\!g}'}{\widetilde m_\Sigma^2\tt\;\!Z}
+ {\tt J_0'} \frac{{\tt Y^{\prime2}\!-\!4~\!\!r}{}_{\!\pi}'}{\tt Z} \,, ~~~ ~~~~
\textsc c_{\;\!\!g}' = \Bigg( {\tt J}_0^{\prime~\!}{\tt r}_{\!\pi}'
- \frac{{\tt C}_\Sigma'{\tt Y'\!+\!C}_p'{\tt r}_{\!\pi}'}{2} \Bigg) \frac{\widetilde m_\Sigma^2}{2} \,, ~~~ ~~~~
{\tt r}{}_{\;\!\!\pi}' = \frac{m_{\pi^0}^2}{\widetilde m_\Sigma^2} \,,
\vspace{1pt} \\ & \displaystyle \!\!
{\tt I_0'} = \frac{\pi|\mbox{\bf\footnotesize P}_{\;\!\!\!\pi}'|}{\widetilde m_\Sigma^{}} \,, ~~~ ~~~~
{\tt J_0'} = \frac{\pi}{4\widetilde m_\Sigma^{}\;\! |\mbox{\bf\textit q}|} {\rm\,\ln}\Bigg(
\frac{ m_{\pi^0}^2 - {\tt2~\!\!E}_\pi' {\tt E}_p + 2|\mbox{\bf\footnotesize P}_{\;\!\!\!\pi}'|
|\mbox{\bf\textit q}| }{ m_{\pi^0}^2 - {\tt2~\!\!E}_\pi' {\tt E}_p
- 2|\mbox{\bf\footnotesize P}_{\;\!\!\!\pi}'| |\mbox{\bf\textit q}| } \Bigg) ,
\\ & \!\! \!\!\!
{\tt E}_\pi'=\sqrt{m_{\pi^0}^2{+}\mbox{\bf\footnotesize P}_{\;\!\!\!\pi}^{\prime2}} \,, ~~~ ~~~~
{\tt E}_p^{}=\sqrt{m_p^2{+}\mbox{\bf\textit q}^2} \,, ~~~ ~~~~
|\mbox{\bf\footnotesize P}_{\;\!\!\!\pi}'| = {\displaystyle\frac{\widetilde m_\Sigma^{}}{2}}
\sqrt{{\tt v_+^2{-}r}{}_{\;\!\!\pi}'}\, \sqrt{{\tt v_-^2{-}r}{}_{\;\!\!\pi}'} \,. \end{array}
\end{align}

These results agree with those derived in ref.\,\cite{He:2005yn} if the nucleons and pions assume their respective isospin-averaged masses, except the common overall factor $m_{\pi^+}^2$ in eq.\,(\ref{imdS}) which originates from the normalization of ${\cal M}_{\Sigma^+\to n\pi^+,\,p\pi^0}$ in eqs.\,(\ref{MSigmaNpi+})-(\ref{MSigmaNpi0}).
The same mass choices were made in refs.\,\cite{Roy:2024hqg,He:2026chn}.
For completeness, as explained in ref.\,\cite{He:2026chn},  employing the data
\,$
\mathscr B_{\Sigma^+n\pi^+}^{\tt exp} = (49.87 \pm 0.30)\%$,\, \,$\alpha_{\Sigma^+n\pi^+}^{\tt exp} = 0.0514 \pm 0.0029$,\, $\mathscr B_{\Sigma^+p\pi^0}^{\tt exp} = (49.79 \pm 0.23)$, and \,$\alpha_{\Sigma^+p\pi^0}^{\tt exp} = -0.9868 \pm 0.0019$\, from refs.\,\cite{ParticleDataGroup:2026mpi,BESIII:2020fqg,BESIII:2023sgt,BESIII:2025jxt,BESIII:2025rgd}, along with the $\Sigma^+$ lifetime and hadron masses from ref.\,\cite{ParticleDataGroup:2026mpi}, we arrive at \,$\mathbb A_{\Sigma^+n\pi^+} = 0.0472 \pm 0.0027$,\, $\mathbb B_{\Sigma^+n\pi^+} = 18.836 \pm 0.064$,\, $\mathbb A_{\Sigma^+p\pi^0} = -1.3853 \pm 0.0079$,\, and \,$\mathbb B_{\Sigma^+p\pi^0} = 11.799 \pm 0.088$,\, which translate into the Im$\,a_\Sigma^{}(0)$ and Im$\,b_\Sigma^{}(0)$ entries in table~\ref{imaimb}.

\bibliography{newrefs}

\end{document}